\documentstyle[12pt]{article}
\input{epsf.tex}
\begin{document}

\begin{center}
{
{\bf A PROFUSION OF BLACK HOLES \\
FROM TWO TO TEN DIMENSIONS}
{\noindent\footnote{\noindent To be published in  the Proceedings of 
the XVII$^{\underline{\rm th}}$ Encontro Nacional de F\'{\i}sica de 
Part\'{\i}culas e Campos, in commemoration of the 30$^{{\underline{\rm th}}}$ 
anniversary of the Sociedade Brasileira de F\'{\i}sica (SBF), 
September 1996, Serra Negra, SP, Brazil.}}
} \\
\vskip 1mm
{\bf Jos\'e P. S. Lemos} \\
\vskip 3mm
{\scriptsize  Departamento de Astrof\'{\i}sica,
	      Observat\' orio Nacional-CNPq,} \\
{\scriptsize  Rua General Jos\'e Cristino 77,
	      20921 Rio de Janeiro, Brazil, \&} \\
{\scriptsize  Departamento de F\'{\i}sica,
	      Instituto Superior T\'ecnico,} \\
{\scriptsize  Av. Rovisco Pais 1, 1096 Lisboa, Portugal.}
\end{center} 
\begin{abstract}
\noindent
Black holes in several dimensions and in several theories are studied 
and discussed. The theories are, general relativity, Kaluza-Klein, 
Brans-Dicke, Lovelock gravity and string theory.  
\end{abstract}
\noindent
{\bf 1. Introduction}

\vskip 2mm

Black hole physics and black holes (BHs) have by now a long and interesting 
history since they were first predicted in 1939 by the prescient work 
of Oppenheimer and Snyder \cite{oppie} following some hints left by 
Zwicky in 1934 \cite{zwickybaade} that neutron stars, stars of 
very high densities and very small radii, could form as the end product 
of a supernova explosion. 

It is not here the place to comment on the development of these ideas, 
but maybe, some would like to know that in the same year, Einstein 
published a paper \cite{einstein39} arguing forcefully that the 
gravitational radius, what we now call the event horizon of a BH, could never  
be surpassed. Einstein was, in a sense, isolated in Princeton, while 
Oppenheimer was on the west coast, the other side of the country, 
commuting with his students 
between Berkeley and Caltech each six months. In Caltech he 
could share ideas with Tolman the great relativist, and Zwicky a 
master of prophesying correctly (although there is no direct sign 
of communication between Zwicky and Oppenheimer). With hindsight, 
it seems that Caltech was the right place to study gravitational 
collapse and predict BH formation. 

It is also relevant to note that 150 years before, dark stars, the 
Newtonian BHs, were predicted by Michell \cite{michel1784}
in Cambridge, an idea that Laplace  followed 12 years later 
\cite{laplace1796}. 
In modern terms Michell's idea can be put in the form: give a mass $M$ of 
an astronomical object; find its radius so that the escape velocity is 
the velocity of light $c$. The answer is $R=\frac{2GM}{c^2}$, where 
$G$ is the gravitational constant. Objects with 
radii below this value are dark stars. 
However, the argument is not strictly valid 
because $c$ does not have a fundamental meaning in Newton's gravity. 
One could detect tachyonic particles emmited from the surface of the 
star, or an observer not placed at infinity, in the neighborhood 
of the star, say, could still see the light coming from the star. 
However entertaining was the dark star idea, it was dropped 
down for one or other reason until 1939, where it appeared in the right 
context, the theory of general relativity. Curiously enough, a good condition 
for the formation of a BH is that the radius of the star obeys Michell 
condition $R=\frac{2GM}{c^2}$, although now $M$ and $R$ have the corresponding 
relativistic meanings and $G$ and $c$ are both fundamental constants. 

So, what is the picture of a star collapsing into a BH? One can best see
it through a spacetime diagram. As the star collapses there is a last ray 
emerging from the center that can reach spatial infinity. This is the 
event horizon, signaling the existence of a BH, see figures 1 and 2.

\vskip 1mm
\centerline{\epsffile{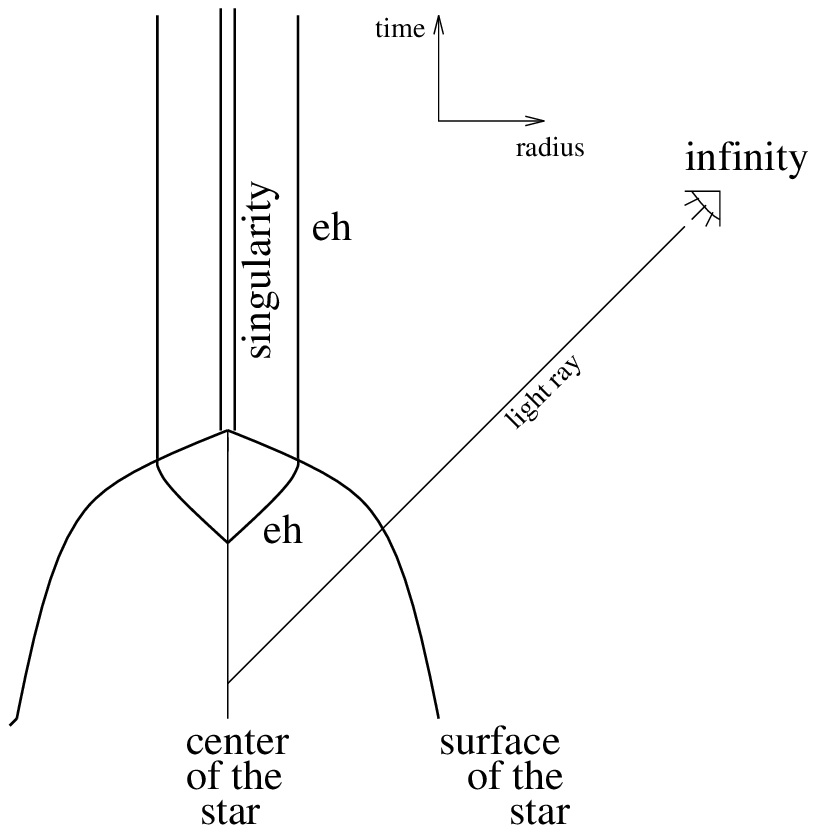}}
\vskip 1mm
{\noindent Figure 1. Eddington-Finkelstein diagram for the collapse of a star, 
(eh = event horizon). A double line in all figures represents a 
polynomial singularity, where curvatures and densities of infinite 
strength are formed. } 

\centerline{\epsffile{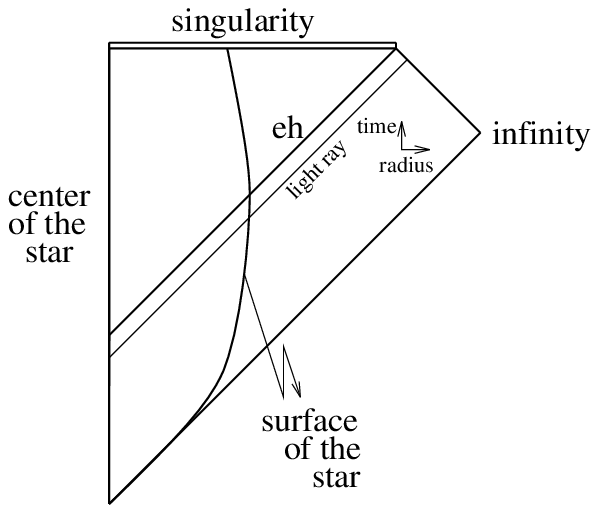}}
\vskip 3mm
{\noindent Figure 2. Penrose diagram for the collapse of the same star 
of figure 1. Light rays move at $\pm 45^{\rm o}$ and each point in the diagram 
represents a 2-sphere. }
\vskip 5mm

When the BH forms there are two distinct but connected regions,  
the inside and the outside of the event horizon, explicitly showing 
that time in relativity is observer dependent. As the matter of the star
continues to collapse inside the event horizon it will form a
singularity where curvatures and densities of infinite strength are
formed and  the usual concept of spacetime is lost. 
Inside the event horizon light is trapped. Light  not only
does not escape to infinity, it cannot escape to the outside of the
BH.  However, to an outside observer the story is different.
As the BH is being formed, the luminosity of the original star decays
exponentially, $L=L_o e^{-\frac{t}{\tau}}$ where the characteristic
time is very short, $\tau=3\sqrt3\frac{GM}{c^3}= 2.6 \rm{x} 10^{-5}
\frac{M}{M_\odot}{\rm s}$, i.e., in a few millionths of a second the star
turns totally black ($M_\odot= {\rm solar}$ ${\rm mass }$).  
In addition, to an outside observer the collapse of
the star results in a BH whose properties are characterized by
three parameters only:  mass, charge and angular momentum. One then
says that BHs have no hair (in fact, they have three hairs).
All the other properties, or `hairs', of the matter of the star 
that formed the BH disappear.  No observation can reveal the nature of the
original star, whether it possessed anti-matter,
or was made of fermions, or bosons, or whether it had any other hairs.

This picture is drastically altered if the collapse produces a 
singularity first, not dressed by an event horizon.  
BHs are well studied and their existence is highly plausible. Naked 
singularities do not enjoy the same status. They are a threat to the 
predicability power of general relativity, and for this reason 
a cosmic censorship conjecture forbidding the existence of such nasty 
objects was formulated \cite{penrose69}. 
There are many theoretical counter-examples to the cosmic censorship 
conjecture \cite{lemosprl92,joshi93}, 
although it is still arguable that these examples cannot 
occur in nature, either because they may be physically unrealistic or 
possibly highly unstable. 
One drawback to the conjecture, often invoked, is that its validity 
implies the impossibility of observing quantum gravity phenomena, 
coming out right from the singularity. 

BHs formed from the collapse of stars can have masses between 
$3-100M_\odot$. There is also the possibility that supermassive stars 
or the core of star clusters collapse to form BHs with masses of 
the order of $1000M_\odot$.  
BHs with much higher masses $10^6-10^9M_\odot$ may form in the center 
of a galaxy via gravitational collapse of a mixture of clusters of stars 
and gas. Primordial BHs with masses ranging up to 
$10^{-19}M_\odot\simeq 10^{14}{\rm g}$, and the radius of a proton 
$10^{-13}{\rm cm}$, could have been formed in the fluctuations of the early 
and very early universe. 

For stellar size objects, the mass is a good indicator to separate BHs 
from neutron stars. If the compact object has a mass 
$M \buildrel>\over\sim 3.5 M_\odot$ then there is no equation of 
state, however stiff, able to support the neutron star (a cold star 
with a radius of $\sim10$Km) against complete collapse. 
There are strong candidates in the sky to stellar BHs,  
the most famous of all is Cygnus X1, a binary system  emiting X-rays and 
harboring a dark compact object with $\sim 16M_\odot$ 
(see  e.g. \cite{lemos96} for a review). 
There are no candidates for BHs with $\sim1000\,M_\odot$ 
(even the existence of supermassive stars is pure theoretical speculation). 
Galactic BHs should inhabit the center of active galactic nuclei, compact 
sources which can shine more than an entire galaxy. In some cases 
like quasars, the nuclei of the galaxy has a brightness equivalent to  
the brightness of several thousands of galaxies, in a region not bigger 
than the solar system. 
In two galaxies with active galactic nuclei the value of the 
central mass points to the existence of a BH: (i) in the 
elliptical galaxy M87 the Hubble Space Telescope measured a rotation 
velocity of $v\sim 550$Km/s for the gas at an orbital radius of 
$60$ light years, which, through 
Kepler's law gives $M= \frac{v^2 R}{2G} \sim 2-3 {\rm x} 10^9M_\odot$; 
(ii)  for the spiral galaxy NGC 4258 Keplerian velocities of 
$\sim 1000$Km/s in an inner orbit of very small radius, $R\sim 0.4$ly,
have been measured through water masers which imply a 
central mass of $M \sim 2 {\rm x}10^7M_\odot$. This work is considered to 
provide the strongest case for a supermassive BH in the center confirming 
the predictions of Lynden-Bell \cite{lynden-bell}, 
(see  \cite{lemos96} for a review). 
All these methods are indirect, and to probe directly the existence of 
a BH one should measure relativistic speeds of 
the matter circulating in the disk very near the event horizon. In addtion, 
when the gravitational antennas are operating we should directly 
detect the formation of BHs either through collapse of a single star, 
or through the merging of binary systems. There is no observational evidence 
for the existence of primordial BHs. 

A quantity that gives some insight to the physical processes occuring 
during the collapse is the average density of the 
collapsing matter $\rho$ when the BH is forming, i.e., when 
$R=\frac{2GM}{c^2}$, yielding 
$\rho =\frac{3c^6}{32\pi G^3}(\frac1M)^2\simeq 1.3 {\rm x}10^{16}
(\frac{M_\odot}{M})^2\frac{\rm gm}{\rm cm^3}$. For a 1$M_\odot$ BH this gives 
a density ten times larger than the nuclear density, whereas for a
$10^8M_\odot$ BH it gives the density of water. This means the larger the 
mass the less uncertain is the physics at the BH formation. Even if 
BHs have not been produced in our cosmos, one could envisage an 
astronomical experiment, by assembling a very large mass in the form 
of dust and let it alone to collapse to form a BH. 
After the matter has passed its own gravitational radius, 
the singularity theorems \cite{penrose65} 
plus theoretical models indicate that the 
density raises to infinity, $\rho\rightarrow\infty$. 
Is it really infinity? In principle there are strong suggestions 
that there is a minimum scale, the Planck scale (constructed from $G$, 
$c$ and Planck's constant $\hbar$), below which the usual 
physical concepts break down. At the Planck scales,  
$R_{\rm pl} = \sqrt{\frac{G \hbar}{c^3}}\simeq 10^{-33} {\rm cm}$ 
and  $M_{\rm pl} = \sqrt{\frac{\hbar c}{G}}\simeq 10^{-5}{\rm gm}$, 
the density of the matter is $\rho \simeq \frac{M}{R_{\rm}^3} =
10^{92} \left(\frac{M}{M_{\rm pl}}\right){\rm gm/cm^3}$.  
At these scales it is expected that the topology of the 
spacetime breaks down in order to accomodate these large masses in 
such a small volume. It is interesting to note that the Planck density 
$\rho_{\rm pl} = \frac{c^5}{G^2\hbar}\simeq 10^{92} {\rm gm/cm^3}$ is the 
density at which a Planck mass turns into a BH,  as well as merging into 
the singular structure of the spacetime. 
General relativity provides an adequate description of BHs that are 
much bigger than the Planck mass. On the other hand for Planckian 
BHs a description in terms of general relativity breaks down and it 
has to be replaced by a quantum theory of gravity.

Even much before the quantum gravity regime starts to be important, 
the BH already presents a quantum mechanical behavior. 
Indeed following hints that 
a BH has an associated entropy and therefore, through the relation 
$S=Q/T$, an associated temperature, Hawking \cite{hawking74}
using quantum field theory on a BH background found that BHs are not 
black but radiate with a blackbody spectrum at a temperature 
$T= \frac{\hbar c^3}{8\pi G k_B}\frac1M \simeq 6 {\rm x}
10^{-8} (\frac{M_\odot}{M}) K$, and have an associated  entropy $S_{\rm BH}$ 
given by $S_{\rm BH} = \frac{k_{\rm B}c^3}{\hbar G}\frac{A}{4}$, where 
$A$ is the area of the BH and $k_{\rm B}$ is the Boltzmann constant. 
Since so many fundamental constants
enter these formulas one can say that quantum mechanics, general relativity 
and thermodynamics must merge together in a unified theory.
For $M\sim 1M_\odot$ one has $T\sim 10^{-7}$K, whereas for a Planckian BH, 
$M\sim 10^{-5}$gm, $T\sim 10^{32}$K. 
An important unsolved problem raised by this thermal evaporation 
is the information paradox, which is the problem of knowing to where 
all the information contained inside the original star has gone after 
the BH has evaporated completely \cite{hawking76,strominger96}.

Classically, BHs are stable objects, however quantum mechanically they are 
unstable. As the BH radiates energy its mass decreases, the temperature 
increases in a runaway process which probably ends in a final explosion. 
Suppose now that instead of neutral BHs one considers a charged non-rotating 
BH. Then, dropping the fundamental constants, 
$T=\frac{1}{2\pi}\frac{\sqrt{M^2-Q^2}}{(M+\sqrt{M^2-Q^2})^2}$. 
If the charge is large enough, $|Q| = M$, then $T=0$ and one could expect 
these objects to be stable. However, vacuum polarization effects will 
discharge the BH itself rapidly. 
There are two ways to stabilize the situation: 
\begin{enumerate}
\item Take a topological charge so that there are no particles 
to radiate \cite{colemanetal1992}. 
\item A charged BH will preferentially radiate away its charge, 
depending on the charge to mass ratio of the particles in the 
theory. If $\frac{q}{m}$ is small most of the radiation will be 
in the form of neutral particles and $Q$ will remain constant. 
Take then that the lightest charged particles are heavy enough 
so that they cannot be created by the BH. This could be done in two 
instances.
\begin{enumerate}
\item For example, suppose that the BH carries magnetic charge 
instead of electric charge. The only way for the BH to loose this charge 
would be via the creation of monopoles. However, if the monopoles are 
heavy enough the probability of decay is heavily suppressed \cite{carter74}. 
\item A variant of this scenario is to suppose that the charge 
arises as a central charge in a supersymmetric algebra. It is known that 
in $N=2$ supergravity the bosonic sector is Einstein-Maxwell theory with 
a Bogomolnyi bound given by $Q\leq M$. 
One can then show that extreme Reissner-Nordstrom solutions 
$|Q|=M$ (which saturate the bound) 
are supersymmetric, in the sense that under a supersymmetric 
operation the metric remains invariant and the fermionic sector 
remains null \cite{gibbonshull1982}. These BHs have zero $T$ and are stable. 
\end{enumerate}
\end{enumerate}

Stable BHs can be considered as solitons of the theory and  
as such belong to the non-perturbative sector and should be put on 
the same foot as the elementary particles of the theory.
To see more directly that the distinction between BHs and elementary 
particles can be blurred, suppose there is an elementary particle with 
a mass greater or equal to the Planck mass. Then its Compton wavelength 
is smaller or equal to its Schwarzschild radius. At these scales it is 
therefore hard to distinguish between what is an elementary particle 
from what is a BH. It is then natural to think of such particles as BHs 
and conversely BHs may be viewed as elementary particles \cite{hawking1970}. 
It is expected that gravity must become the dominant field at 
the quantum Planck scale $10^{-33}$cm, which as we have said represents 
the minimum scale at which spacetime can be considered smooth. 
BHs, viewed as elementary particles,  are the objects to test this scale, 
through Hawking radiation. Imagine the following futuristic experiment:
two incoming particles in a huge accelerator are set to collide 
face-on, such that, a center of mass energy of
$\sim 10^{19}$Gev is produced. Then, one might form a Planckian BH which will 
evaporate quickly in a burst, allowing us to study the physics at 
the Planck scale. One might think that by increasing the energy 
the study of sub-Planckian scales would follow. However, this is 
not the case, since one would produce a BH with larger mass, which 
would decay slowly.

From all this one can see that quantum gravity plays an essential role 
in every theory of extremely strong gravitational fields such as 
BHs and singularities. One could think of reconciling general relativity 
with quantum mechanics, but it is known that general relativity is 
perturbatively unrenormalizable which is taken at face value by many 
people as an indication that the quantum theory does not exist. 
At present, the best candidate to a consistent theory of quantum gravity is 
string theory, a theory remarkable in some respects. The idea of string 
theory is to use strings as fundamental entities and treat its vibrations 
as manifestations of the physical world, as fields, particles, etc. 
The string action plus some rules (like preservation of conformal invariance 
at the quantum level) place strong restrictions 
on the possible theories and on the spacetime itself. For instance, string 
theories treat the dimension of spacetime as a parameter to be settled
by the theory. For the pure bosonic string theory (inconsistent at 
the quantum level), the dimension is $D=26$, 
while $D=10$ for the four consistent supersymmetric string theories 
which seem to belong to a $D=11$ M--theory \cite{townsend96,witten95} 
or even a $D=12$ F--theory \cite{vafa}. 
Although apparently incorrect, these dimensions can, in principle, be 
dynamically compactified into the $D=4$ dimensions actually observed 
in our universe. Superstring theories can also be formulated 
in any dimension $D\leq 10$, with the left $10-D$ 
dimensions treated as being compactified somehow \cite{greenschwarzwitten}.
A remarkable feature of the theory is the presence of a bewildering 
variety of BH solutions in any dimension from 2 to 10. The study of 
BH solutions in $D\geq4$ dimensions is not new \cite{tangherlini}, 
although string theory has made an important impact in their development 
in higher as well as lower (2 and 3) dimensions.  
Besides string theory, BHs in different dimensions also appear in 
theories like  general relativity, Kaluza-Klein theory, 
Brans-Dicke theory, Lovelock gravity and in their corresponding 
supersymmetric versions. In the subsequent sections we will discuss some 
of these solutions and some of their properties. 
Some discussion of sections 2 and 5 is patterned along the lines of 
\cite{horowitzreview}, and part of section 3 follows \cite{carlipCQG}.

\vskip 5mm

\noindent
{\bf 2. BHs in ${\bf 4D}$} 
\vskip 2mm

Let us start with general relativity in 4$D$, i.e., Einstein-Maxwell 
theory, characterized by the action 
\begin{equation}
S=\frac{1}{16\pi G}\int d^4 x \sqrt{-g}(R-F^2) \, ,
                         \label{eq:2.1}
\end{equation}
where $g$ and $R$ are the determinant of the metric and the curvature 
scalar, respectively, and $F^2=F_{\mu\nu} F^{\mu\nu}$, where 
$F_{\mu\nu}$ is the Maxwell tensor ($c=1$). 
Uncharged static BHs are described by the Schwarzschild solution 
\begin{equation}
ds^2 = -(1-\frac{2M}{r}) dt^2 + \frac{dr^2}{1-\frac{2M}{r}} + 
r^2 d{\Omega_2}^2 \\ ,
                         \label{eq:2.2}
\end{equation}
where $d{\Omega_2}^2$ is the line element of the 2-sphere, 
$M$ is the mass of the BH, and we have put $G=1$. 
The causal structure is conveniently 
described by the  Penrose diagram of figure 3, where light rays move at 
$\pm 45^{\rm o}$ and each point in the diagram represents a 2-sphere. The 
event horizon is located at $r=2M$ (where $g^{rr}=0$). 
\vskip5mm
\centerline{\epsffile{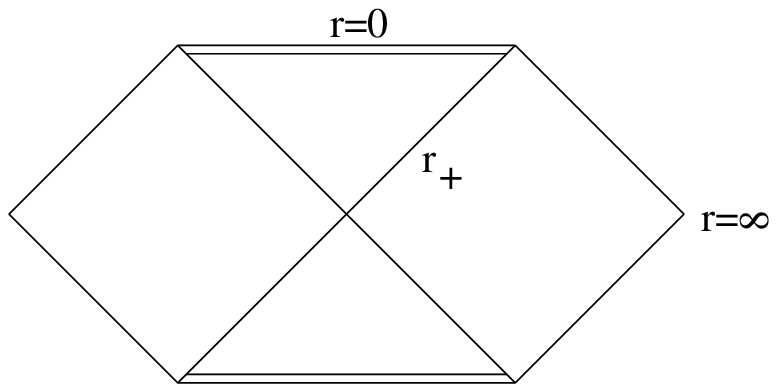}}
\vskip 3mm
\centerline{Figure 3. Penrose diagram for the Schwarzschild solution.}
\vskip 5mm

A charged static BH in general relativity is described by the 
Reissner-Nordstrom solution, 
\begin{equation}
ds^2 = -(1 - \frac{2M}{r} + \frac{Q^2}{r^2}) dt^2 + 
\frac{dr^2}{1-\frac{2M}{r} + \frac{Q^2}{r^2}} + 
r^2 d{\Omega_2}^2 \\ ,
                         \label{eq:2.3}
\end{equation}
where $Q$ is the charge, $F_{rt}=\frac{Q}{r^2}$ for electric $Q$, 
and $F_{\theta\phi}=Q\sin\theta$ for magnetic $Q$. The causal structure 
is richer now. There are three distinct cases depending on the charge 
to mass ratio. For $0<|Q|<M$  there are two horizons (the event 
and the Cauchy horizon)  given by the roots of $g^{rr}=0$, $r_{\pm}$. 
The Penrose diagram is sketched in figure 4. 
For an extreme BH, $|Q|=M$, the two horizons merge in one, and 
for $Q>M$ the singularity is timelike and naked.

The Hawking temperature of static BHs can be calculated in 
several ways. The original calculation involves the analysis of 
quantum matter fields in the BH background \cite{hawking75}. 
A cleaner calculation is achieved by  analitically continuing 
the metric in time $t$ and requiring that the resulting 
Riemannian space be non-singular. This implies a periodic identification 
in imaginary time with the temperature being equal to the inverse of 
the period \cite{hawking79}. One can then show that this BH instanton is 
related to a real BH in thermal equilibrium with radiation. 
As mentioned, for the Reissner-Nordstrom BH 
$T=\frac{1}{2\pi}\frac{\sqrt{M^2-Q^2}}{(M+\sqrt{M^2-Q^2})^2}$, 
which for $Q=0$ yields the familiar $T=\frac{1}{8\pi M}$. 

\vskip5mm
\centerline{\epsffile{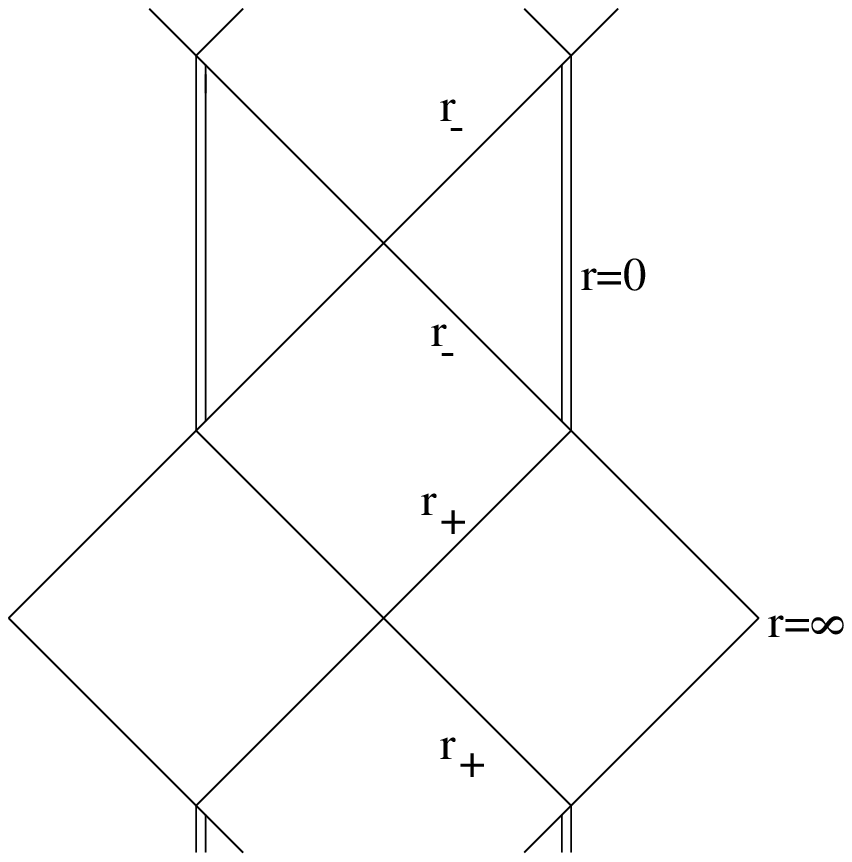}}
\vskip 3mm
\centerline{Figure 4. Penrose diagram for the Reissner-Nordstrom solution.}
\vskip 5mm

Near singularities general relativity should be replaced by a quantum 
theory. String theory is a consistent theory that may give some clues 
at the Planckian scales. This raises the question of whether BHs in 
string theory are different from BHs in general relativity. We will see 
that these two theories give distinct BHs. Due to the existence 
of dilaton, axion and other fields in string theory there are 
even BHs without singularities. There are also solutions describing 
one-, two-, and p-dimensional objects surrounded by event horizons, i.e., 
black strings, black membranes and black p-branes. We will also show in the 
next section that general relativity also possesses these type of objects, 
a feature not known untill recently \cite{lemosplb95,lemoscqg95}.  

Without further details for the time being, let us consider the low 
energy action to heterotic string theory 
\cite{greenschwarzwitten,horowitzreview}
\begin{equation}
S= \frac{1}{4\pi}
\int d^{D}x\sqrt{-g} e^{-2\phi} \left[ R - 2\Lambda + 4(\nabla \phi)^2
- F^2 - \frac{1}{12} H^2 
\right] \, , 
                         \label{eq:2.4}
\end{equation}
where the new fields are the dilaton scalar field $\phi$, 
and the 3-form field $H_{\mu\nu\rho}$, such that 
$H^2 = H_{\mu\nu\rho} H^{\mu\nu\rho}$ and defined by 
$H=dB-A\times F$ where $B_{\mu\nu}$ is the axion  
2-form potential and $A_{\mu}$ is the vector potential that defines 
the $U(1)$ Maxwell field, $F=dA$. These fields arise naturally in string 
theory. The cosmological constant $\Lambda$ is set by the internal 
consistency of the theory and related to the dimension $D$ of the spacetime
and the central charge of a possible internal conformal field theory. 
The constant factor $\frac{1}{4\pi}$ in front of the integral in the action 
(\ref{eq:2.4}) is somewhat arbitrary. This arbitrariness will remain 
throughout this article, although without loss of precision, since we are 
dealing mostly with classical results. 

To have a full theory and not only the low energy action (\ref{eq:2.4}) 
one would have to add higher order correction terms 
$R^2$, $R^3$, $F^4$, etc. All the higher order terms are 
important for studying BHs of Planckian size and the spacetime singularities. 
However, using (\ref{eq:2.4}) one can study the properties of larger BHs 
away from the singularities. 
For $D=4$ and in a background where $\Lambda=0=H$ the action 
simplifies to
\begin{equation}
S= \frac{1}{4\pi}
\int d^4x\sqrt{-g} e^{-2\phi} \left[ R + 4(\nabla \phi)^2
- F^2 \right] \, .
                         \label{eq:2.5}
\end{equation}  
Note that $\phi$ plays the role of a coupling constant, since comparing 
(\ref{eq:2.1}) and (\ref{eq:2.5}) roughly one has 
$G\sim e^{2\phi}\equiv g_s$, where $g_s$ is the string coupling constant. 
In order to directly compare with the Einstein-Maxwell action one 
rescales the string metric $g_{\mu\nu}$ 
(which is the metric seen by the strings) 
to the Einstein metric $g^E_{\mu\nu}\equiv e^{-2\phi} g_{\mu\nu}$ (the metric 
that puts the string action in an Einstein form) to have the action,
\begin{equation}
S_E= \frac{1}{4\pi}
\int d^4x\sqrt{-g_E}\left[ R_E + 4(\nabla \phi)^2
- e^{-2\phi}F^2 \right] \, . 
                         \label{eq:2.6}
\end{equation}  
For $F=0$, i.e., uncharged solutions,  one deduces 
from (\ref{eq:2.6}) and the no-hair theorems \cite{crusciel} that  uncharged 
BHs in the low energy string action are the same as the Schwarzschild BH 
of general relativity. On the other hand, for $F\ne0$ and $\phi\ne0$ 
the charged BHs in 
string theory are different from the Reissner-Nordstrom BHs. This could 
give a low energy test of string theory: if string theory is the correct 
one then charged BHs are not described by the Reissner-Nordstrom metric 
but instead by the solution \cite{gibbons82,maeda88,garfinkle91}
\begin{eqnarray}
& ds^2=-(1-\frac{2m}{\overline r}) (1+\frac{2m\sinh^2\alpha}{\overline r}) dt^2 
+ \frac{d{\overline r}^2}{1-\frac{2m}{\overline r}} + {\overline r}^2 
d{\Omega_2}^2 \,  &
\nonumber \\ 
& e^{-2\phi} = 1 + \frac{2m\sinh^2\alpha}{\overline r} \, 
A_t = - 
\frac{m\sinh2\alpha}{\sqrt2\left[ {\overline r} + 2m\sinh\alpha\right]}
\, , & 
                         \label{eq:2.78}
\end{eqnarray}
where the mass and charge are given by 
$M=m\cosh^2\alpha$, $Q=\sqrt2 m \sinh2\alpha$. 
For ${\overline r} = 2m$ there is an event horizon whereas for 
${\overline r} = 0$ there is a singularity. At the singularity 
$g_s = e^{2\phi}\rightarrow 0$ which might mean that in the full 
string theory, the string coupling remains negligible and 
quantum effects are suppressed. 
To compare with general relativity we then do the conformal 
rescaling mentioned above  ($ds_E^2=e^{-2\phi}ds^2$) and obtain 
\begin{eqnarray}
& ds_E^2 = -(1-\frac{2M}{r}) dt^2 + \frac{dr^2}{1-\frac{2M}{r}} + 
r(r-\frac{Q^2}{r}) d{\Omega_2}^2 \, , &
\nonumber \\
& e^{2\phi} = 1-\frac{Q^2}{M r} \, , F_{rt} = \frac{Q}{r^2} \, ,
&
                         \label{eq:2.9}
\end{eqnarray}  
where for convenience we have defined $r={\overline r} + \frac{Q^2}{M}$. 
The charged string metric  is identical to Schwarzschild in the $r-t$ plane 
(same Penrose diagram as in figure 1), however the spheres have smaller 
radii. There is the extremal limit $|Q|=M$ given by the diagram of 
figure 5. For $|Q|>M$ the singularity is naked. The string metric 
(\ref{eq:2.78}) has the same  corresponding 
Penrose diagrams since these diagrams are 
unaltered by conformal transformations.  

\vskip5mm
\centerline{\epsffile{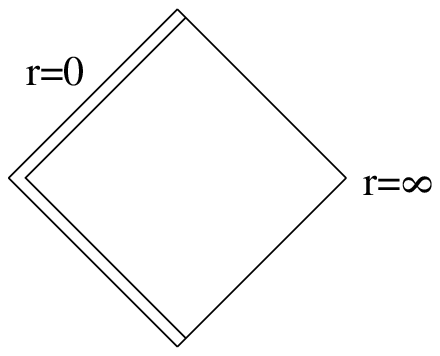}}
\vskip 3mm
{\noindent Figure 5. Penrose diagram for the charged extreme BH in string 
theory. The singularity is null, or in other words, the event horizon 
is singular.}  
\vskip 5mm

What about magnetic BHs? We have seen that in general relativity, 
electric and magnetic BHs have the same metric, i.e., neutral particles 
do not distinguish the two types of BHs. In string theory one can find 
magnetic BHs by performing an S-duality (or strong-weak) transformation, 
which transforms weak coupling into strong coupling and vice-versa. 
The transformation is \cite{horowitzreview} 
\begin{equation}
F\rightarrow {\tilde F} \, , \,\,  \phi\rightarrow -\phi \, ,  
\,\, g_E\rightarrow g_E
                         \label{eq:2.10}
\end{equation} 
where ${\tilde F}$ is the dual of $F$, 
${\tilde F}_{\mu\nu}=\frac12 e^{-2\phi} 
{\epsilon_{\mu\nu}}^{\alpha\beta} F_{\alpha\beta}$, 
transforming electric into magnetic charge. 
Since the Einstein metric is unchanged the Penrose diagrams for 
magnetic BHs are identical to 
the Penrose diagrams for electric BHs. 
In terms of the string metric we have
\begin{eqnarray}
& ds^2 = - \frac{1-\frac{2M}{r}}{1-\frac{Q^2}{Mr}} dt^2
+ \frac{dr^2}{(1-\frac{2M}{r})(1-\frac{Q^2}{Mr})}
+r^2 d{\Omega_2}^2 \, &
\nonumber \\
&  e^{-2\phi} = 1 - \frac{Q^2}{Mr} \, , \,\,\,F_{\theta\phi} = Q\sin\theta  
&   
                \label{eq:2.11}
\end{eqnarray} 
The singularity happens at a finite area, 
when $r=\frac{Q^2}{M}$. The extremal limit is given by $Q^2=2M^2$, 
for which the temperature is zero. On the other hand for the non-extreme 
BH given in equation (\ref{eq:2.11}), the temperature is 
$T = \frac{1}{8\pi M}$, 
independent of the charge. This means that the BH radiates past beyond 
the extremal limit, indicating in turn that the semi-classical approximation 
for the calculation of the temperature breaks down. 

We have only mentioned non-rotating BHs. In string theory, uncharged 
rotating BHs have the same metric as Kerr BHs. However the charged 
rotating BHs are different \cite{sen92}.

\vskip 5mm

\noindent
{\bf 3. BHs in ${\bf 3D}$} 
\vskip 2mm

It is now known that 3$D$ general relativity is important to study 
as it provides a bedtest for 4$D$ and higher $D$ theories 
\cite{Deseretal82,achutownsen88,witten88}. Two features in 
3$D$ general relativity are relevant: (i) the theory has no Newtonian 
limit (it is still an open question which 3$D$ theory has a Newtonian 
limit), (ii) there are no propagating degrees of freedom, which means 
that in vacuum, outside matter, spacetime is locally flat, anti-de Sitter 
or de Sitter depending on the value of the cosmological constant,  
$\Lambda=0$, $\Lambda<0$, and $\Lambda>0$, respectively. 
Due to this simplicity and lack of structure it can be thought that 
there is no interesting object emerging from the theory. 
Surprisingly, from the action 
\begin{equation}
S=\frac{1}{2\pi}\int d^3 x \sqrt{-g}(R-2\Lambda) \, .
                         \label{eq:3.1}
\end{equation}
and its equations of motion, Ba\~ nados, Teitelboim and Zanelli 
\cite{banadosetal92} found a 3$D$ rotating BH metric 
known as the BTZ BH,  given by 
\begin{equation}
ds^2 = - (\frac{r^2}{l^2} - M + \frac{J^2}{4r^2}) dt^2
+ \frac{dr^2}{\frac{r^2}{l^2} - M + \frac{J^2}{4r^2}} 
+ r^2 (d\varphi - \frac{J}{2 r^2} dt) ^2 \, ,
                         \label{eq:3.2}
\end{equation}
where $l^2\equiv - \frac1\Lambda$, $J$ is the angular momentum, and here 
$G\equiv\frac18$. For $|J|<Ml$ there are two horizons 
$r_\pm$ given by the zeros of $g^{rr}$. There are also ergoregions for 
$r_+ < r < r_{\rm erg}$ where particles and observers are dragged along 
certain trajectories. In the extremal case, $|J| = Ml$, the two horizons 
merge. For $J=0$ the BH is static.  The rotating case resembles 
in many aspects the Kerr metric and the non-rotating case the Schwarzschild 
solution, although there are no polynomial singularities,
only  (milder) causal singularities. The maximal analytical extension 
of the static and rotating BHs are given in the Penrose diagrams of 
figures 6 and 7. 

\vskip5mm
\centerline{\epsffile{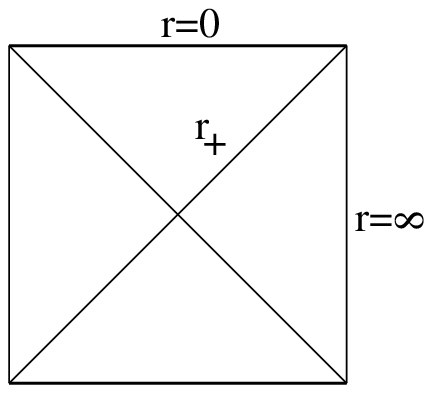}}
\vskip 3mm
{\noindent Figure 6. Penrose diagram for the 3D static BH. The line $r=0$ in 
this figure and in figure 7 is a milder causal (not polynomial) 
singularity. Spacetime is asymptotically anti-de Sitter. }
\vskip 5mm

Besides the BH solution, 3$D$ general relativity with $\Lambda<0$ also 
has the anti-de Sitter (ADS) spacetime as a vacuum solution 
with metric given by
\begin{equation}
ds^2 = - (\frac{r^2}{l^2}+1) dt^2 
+ \frac{dr^2}{\frac{r^2}{l^2}+1} + r^2 d\varphi^2 \, .
                         \label{eq:3.3}
\end{equation}
We note that for $r\rightarrow\infty$ the BH solution (\ref{eq:3.2}) 
is asymptotically ADS. 
Asymptotically ADS solutions and ADS spacetime itself are interesting 
to study for various reasons: (i) theories of extended 
supergravity in which some group, like $O(N)$, 
is gauged have ADS as a vacuum state, and  
(ii) there exists a positive energy theorem, i.e., it is possible to give 
Witten's proof of the positive mass theorem of Schoen and Yau to 
asymptotically ADS spacetimes, implying in turn that asymptotically 
ADS solutions are stable. 

Now, in 3$D$ there is the relation  
${R^{ab}}_{cd} = \epsilon^{abe}\epsilon_{cdf} {G^e}_f$. 
Therefore, a solution of $G_{ab}=0$ is flat, and a solution 
of $G_{ab} = - \Lambda g_{ab}$ has constant curvature.
Since the BH metric and the ADS solution have both constant curvature, 
one concludes that patches in the BH spacetime have an isometric 
neighborhood  to the ADS spacetime and the BH spacetime can be defined 
by a collection of such neighborhoods. Indeed, it was shown in 
\cite{bhtz93} that 
the BH can be represented as a quotient space of the universal covering 
of ADS, $\tilde{\rm ADS}$, by some group of isometries, which provides a 
powerful mathematical tool in examining the BH spacetime. 

\vskip5mm
\centerline{\epsffile{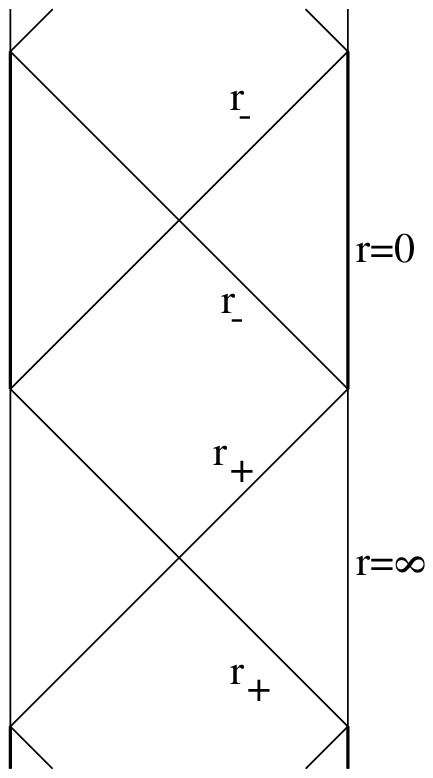}}
\vskip 3mm
\centerline{Figure 7.  Penrose diagram for the 3D rotating BH.}
\vskip 5mm

3$D$ ADS spacetime can be obtained from the plane $R^4$ with two time 
and two space coordinates $(X_1, X_2, T_1, T_2)$ (we follow 
\cite{carlipCQG} here). The ADS metric is then the 
induced metric taken from the 4$D$ flat metric, 
\begin{equation} 
ds^2 = -d{T_1}^2 - d{T_2}^2 + d{X_1}^2 + d{X_1}^2 \, ,
                         \label{eq:3.4}
\end{equation}
restricted to the hyperboloid
\begin{equation}
{X_1}^2 - {T_1}^2 + {X_2}^2 - {T_2}^2 = - l^2 \, .
                         \label{eq:3.5}
\end{equation}
From (\ref{eq:3.4}) and (\ref{eq:3.5}) the isometry group is $SO(2,2)$, 
of course. One can go further and combine $(X_1,X_2,T_1,T_2)$ in a 
$2\times2$ matrix, 
\begin{eqnarray}
{\underline{\underline X}} = 
\left(
\begin{array}{clcr}
   T_{1}+X_{1} & T_{2}+X_{2} \\
 -T_{2}+X_{2} & T_{1}-X_{1}
                        \end{array}
\right)  \label{eq:3.6}
\end{eqnarray}
with ${\rm det} |{\underline{\underline X}}| = 1$ and 
${\underline{\underline X}} \in SL(2,R)$. Here, the isometries can be 
represented as elements of the group 
$SL(2,R){\rm x}SL(2,R)/ Z_2 \approx SO(2,2)$, with each $SL(2,R)$ acting 
by left and right multiplication, such that 
${\underline{\underline X'}} = \rho_L {\underline{\underline X}} \rho_R$, 
with $(\rho_L,\rho_R) \sim (-\rho_L,\rho_R)$. 

Now, given $\tilde{\rm ADS}$ spacetime one may cover it using three 
different regions parametrized by $(r,t,\varphi)$ with 
$0\leq r < \infty$, $-\infty<t<\infty$, and $-\infty<\varphi<\infty$. 
For instance, in the region $r\geq r_+$ we have 
$X_1 = l \sqrt{\alpha(r)} \sinh(\frac{r_+}{l}\varphi - \frac{r_-}{l^2}t)$, 
$T_1 = l \sqrt{\alpha(r)} \cosh(\frac{r_+}{l}\varphi - \frac{r_-}{l^2}t)$, 
$X_2 = l \sqrt{\alpha(r)-1} \cosh(\frac{r_+}{l^2}t - \frac{r_-}{l}\varphi)$, 
and 
$T_2 = l \sqrt{\alpha(r)-1} \sinh(\frac{r_+}{l^2}t - \frac{r}{l}\varphi)$, 
where, $\alpha(r) = \frac{r^2-r_-^2}{r_+^2-r_-^2}$. 
This corresponds to give region I of the Penrose diagram in figure 7. 
Analogous transformations can be given to the regions $r_-< r < r_+$ and 
$0< r < r_-$, i.e., to regions II and III of the figure 7. By repeating 
these regions ad infinitum one covers the entire ADS spacetime. 
One can pick up $X_1,T_1,X_2,T_2$ from these transformations, 
put back in the induced metric  (\ref{eq:3.4})--(\ref{eq:3.5}), and recover 
the form of the BH metric (\ref{eq:3.2}). 
However, note that this is not the BH spacetime 
since $\varphi$ ranges from $-\infty$ to $+\infty$. To make $\varphi$
an angular variable one has to indentify $\varphi$ with $\varphi+2\pi$. 
In this construction it is easy to see that such an identification is an 
isometry of ADS, in fact it is a boost in the $X_1-T_1$ and $X_2-T_2$ 
planes. Indeed, it leads to, 
$X_1\rightarrow {X_1}' = (\cosh \frac{2\pi r_+}{l}) X_1 + 
(\sinh \frac{2\pi r_+}{l}) T_1$, 
$T_1\rightarrow {T_1}' = (\sinh \frac{2\pi r_+}{l}) X_1 + 
(\cosh \frac{2\pi r_+}{l}) T_1$, and analogously for $X_2$ and $T_2$. 
This corresponds in the $SL(2,R)$ formulation 
to an element $(\rho_L,\rho_R)$ given by 
$\rho_L= {\rm diag}
\left(e^{\pi(\frac{r_+-r_-}{l})},e^{-\pi(\frac{r_+-r_-}{l})}\right)$, 
$\rho_R= {\rm diag}
\left(e^{\pi(\frac{r_++r_-}{l})},e^{-\pi(\frac{r_++r_-}{l})}\right)$. 
The BTZ BH may then be viewed as a group manifold given by the quotient 
space $\tilde{\rm ADS}/P$, where $P$ denotes the group generated by 
$(\rho_L,\rho_R)$. 

This formulation has great advantages: 
the ADS spacetime is an extremely simple manifold and if one makes 
appropriate global identifications one finds a 3$D$ BH which 
has inherit its own complex structure . The implications are many: 
(i) one can compute the Green functions in the ADS spacetime and then make 
a direct connection to the BH; (ii) one can find Killing spinors 
fairly easily, which provides an identification of the existence of 
supersymmetry; if the BH is embeded in a supergravity theory with vanishing 
gravitino field, then the existence of Killing spinors leave the metric 
and gravitinos invariant. It was found that 
Killing spinors exist for extreme BHs only 
\cite{henneauxcousaert93}; (iii) the temperature of the BH is 
$T=\frac{r_+^2 - r_-^2}{2\pi r_+ l^2}$, which for zero rotation yields, 
$T=\frac{\sqrt M}{2\pi l}$ and an entropy $S= 4\pi r_+$. 
Unfortunately, this does not help in solving the long standing problem in 
4$D$, to know whether or not the BH evaporates completely, since in $3D$
$T\rightarrow0$ as $M\rightarrow0$; (iv) on the other hand, one can 
show that the $3D$ BH forms from gravitational collapse 
of $3D$ matter, as in the 
$4D$ case \cite{mannross}; (v) $4D$ gravity can be written in a first order 
formalism as a Chern-Simons theory. Viewing the BH as an ADS space with 
proper identifications helps in the study of the holonomies 
(see \cite{carlipCQG} for a complete list of references). 

Another important result, is that the 3$D$ BH we have been discussing  
is also a solution of 3$D$ string theory \cite{horowitzwelch93,kaloper93}. 
Using the action (\ref{eq:2.4}) 
with $D=3$, $\phi=0$ and $H_{\mu\nu\rho} = \frac2l \epsilon_{\mu\nu\rho}$ 
one obtains the same 3$D$ BH. This displays the 
versatility of string theory. One can also find a black string solution by 
applying a duality transformation. We have already seen the S-duality 
at work. There is another well known symmetry of string theory that maps 
any solution with a translational symmetry of the low-energy action into 
another solution. This symmetry is usually called T-duality or 
target-duality. 
Given a target-space solution $(g_{\mu\nu},B_{\mu\nu},\phi)$ which 
is independent of one coordinate, like $\varphi$ in 
the BH solution, then there is another 
solution $({\tilde g}_{\mu\nu},{\tilde B}_{\mu\nu},\tilde{\phi})$ related 
to the previous one by a T-duality \cite{horowitzreview}. The T-dual solution 
for the  $3D$ BH is a black string. 

What else can we do with the 3$D$ BH? It can be embedded in 4$D$ general 
relativity \cite{kaloper94,lemoszanchin961}. One takes the product of the 
BTZ BH with the real line $R$, with  metric 
$ds^2 = ds_{\rm BTZ}^2 + dz^2$, and imposes that it satisfies the 4D 
Einstein equations derived from the action 
$S = \frac{1}{16\pi}\int d^4x \sqrt{-g} \left[(R-2\Lambda) + 
{\cal L}_{\rm matter}\right]$. By 
suitably chosing the energy-momentum tensor 
$T_{\mu\nu} \equiv -\frac{2}{\sqrt{-g}}
\frac{\delta{\cal L_{\rm matter}}}{\delta g_{\mu\nu}}$ 
one finds that the $3D$ BH can be 
converted into a black string in $4D$ general relativity. The idea is analogous 
to the well-known result that point particles in $3D$ are related to 
straight infinite strings in $4D$. 

There is yet a different solution which relates vacuum black strings 
in $4D$ general relativity with $3D$ BHs of a dilaton-gravity theory. 
Starting with the Einstein-Maxwell action 
$S = \frac{1}{16\pi}\int d^4x \sqrt{-g} (R-2\Lambda-F^2)$ one imposes 
the existence of a Killing vector such that the metric can be written 
in the form $ds^2 = g_{ab}^{(3)} dx^adx^b + e^{-4\phi}dz^2$, where 
$a,b=1,2,3$ and $g_{ab}$, and $\phi$ are functions of $x^a$. Then by 
dimensional reduction one obtains a dilaton-gravity action, 
$S = \frac{1}{16\pi}
\int d^3x \sqrt{-g} e^{-2\phi} (R-2\Lambda-F^2)$. It is 
then easy to relate $4D$ and $3D$ solutions. In $4D$ general relativity 
there is a black string solution, with charge and rotation, 
given by \cite{lemoszanchin962} 
\begin{eqnarray}
& ds^{2} = -\left(\alpha^2r^2 -\frac{4M(1-\frac{a^2\alpha^2}{2})}{\alpha r} + 
\frac{4Q^2}{\alpha^2r^2}\right) dt^2 + &
\nonumber \\
&-\frac{4aM\sqrt{1-\frac{a^2\alpha^2}{2}}}{\alpha r}\left(1-
\frac{Q^2}{M(1-\frac{a^2\alpha^2}{2})\alpha r}\right) 2dt d\varphi +&
\nonumber \\
&+ \left(\alpha^2r^2 -\frac{4M(1-\frac{3}{2}a^2\alpha^2)}{\alpha r} + 
\frac{4Q^2}{\alpha^2r^2} 
\frac{(1-\frac{3}{2}a^2\alpha^2)}{(1-\frac{a^2\alpha^2}{2})}
\right)^{-1} dr^2 +&
\nonumber \\
&+ \left[r^2 + \frac{4Ma^2}{\alpha r}\left(1-
\frac{Q^2}{(1-\frac{a^2\alpha^2}{2})M\alpha r}\right)\right] d\varphi^2 + 
\alpha^2r^2 dz^2\, ,&
                             \label{eq:3.7}
\end{eqnarray}
where here $\alpha\equiv-\frac{1}{3}\Lambda$, 
$M$ and $Q$ are the mass and charge, respectively, and 
$a$ is related to the angular momentum $J$ via 
$J=\frac{3}{2}aM\sqrt{1-\frac{a^2\alpha^2}{2}}$, with $0\leq\alpha a\leq 1$. 
This solution has many similarities with the Kerr-Newman BH. For instance,  
the causal structure for the non-extreme BH, i.e., 
$0< a^2\alpha^2 < \frac23 - \frac{128}{81} 
\frac{Q^6}{M^4(1-\frac12a^2\alpha^2)^3}$, is given by the 
Penrose diagram of figure 7, with $r=0$ being now a polynomial 
singularity. 
However, unlike the Kerr-Newman BH, 
the topology of the horizon is cylindrical or toroidal, rather than 
spherical, violating Hawking's theorem \cite{hawkingellis} due to the 
presence of a negative $\Lambda$. It also has implications on the 
hoop conjecture \cite{thorne72}: gravitational collapse in such a background 
can generate a black string even if one is not able to pass a hoop of 
given circunference through the matter. If there is no charge then the 
causal structure changes drastically, resembling the Schwharzschild-ADS 
BH rather than the Kerr BH \cite{lemosplb95}. 

The $3D$ BH generated through dimensional reduction of 
$4D$ general relativity, has a dilaton 
in addition to the metric and Maxwell fields. A study to put these black 
solutions in a supersymmetric context is being carried \cite{lemosmoniz97}. 
Generalizations of the $3D$ action to a Brans-Dicke type of action, 
given by $S = \frac{1}{2\pi}\int d^3x \sqrt{-g} e^{-2\phi} 
(R+4\omega (\nabla\phi)^2 -2\Lambda)$ also yield static and stationary BH 
solutions \cite{sakleberlemos96,salemos95,kchan}. 
Using a metric with two Killing vectors, one can find black membranes 
in general relativity, related through dimensional reduction to $2D$ dilatonic 
BHs. This is a matter for the end of the next section.

\vskip 5mm

\noindent
{\bf 4. BHs in ${\bf 2D}$} 
\vskip 2mm

To analyse BHs in $2D$ we first return to string theory.  In $2D$ there 
is less freedom for dynamics, for obvious reasons.  For instance, for a 
compact orientable $2D$ manifold of genus $g$ (e.g., sphere $g=0$, 
torus $g=1$, etc), the Einstein-Hilbert action, $\frac{1}{2\pi}\int d^2x 
\sqrt{-g} R=2(1-g)$, is the Euler characteristic of space, a topological 
invariant with no dynamics.  Therefore, if one wants to go further in 
$2D$ one has to find a different action.  An interesting action is 
provided by string theory. For understanding the appearance of BHs in $2D$ 
string theory is now important to introduce some basic concepts of the 
theory itself. In string theory one has to distinguish the world-sheet 
action for the string from the target-space or spacetime action for the 
usual spacetime fields. The latter follows from the former upon 
imposing certain restrictions related to renormalization procedures. 
(In particle theory there is also such a distinction but the respective 
actions are not inter-related a priori.) 
The propagation of strings in a generic curved spacetime is described 
by the Polyakov action 
\begin{equation}
S = \frac{1}{4\pi\alpha'}
\int d^2\sigma \sqrt{h} h^{\alpha\beta} \nabla_{\alpha} 
x^\mu  \nabla_{\beta} x^\nu g_{\mu\nu} \, ,
                         \label{eq:4.1}
\end{equation}  
where $h^{\alpha\beta}$ is the world-sheet metric of the string, 
$x^\mu$ are the 
spacetime (or target-space) coordinates, $g_{\mu\nu}$ is the metric 
of the background, and $\alpha'$ is the string coupling 
constant (see figure 8). 
Such an action is also called a non-linear sigma model. 
It is invariant under reparametrizations of the string world-sheet 
$\sigma\rightarrow\sigma'$ and moreover, is conformal invariant 
(i.e, local scale invariant), 
$h_{\alpha\beta}\rightarrow \Omega^2 h_{\alpha\beta}$. 
In principle, one should also include in the action, besides the graviton, 
the other massless states or fields of the (closed) bosonic string, namely, 
the antisymmetric tensor $B_{\mu\nu}$ and the dilaton $\phi$  
(see \cite{greenschwarzwitten} also for the inclusion of fermionic fields 
and supersymmetry). The bosonic world-sheet action or $\sigma-$model is then,  
\begin{eqnarray}
& S =  \frac{1}{4\pi\alpha'}
\int d^2\sigma \sqrt{h} h^{\alpha\beta} \nabla_{\alpha} 
x^\mu  \nabla_{\beta} x^\nu g_{\mu\nu}(x) 
& \nonumber \\
& -\frac{1}{4\pi\alpha'} \int d^2\sigma \epsilon^{\alpha\beta}
\nabla_{\alpha} x^\mu  \nabla_{\beta} x^\nu B_{\mu\nu}(x) 
+ \frac{1}{4\pi} \int d^2\sigma \sqrt{h} R_{\rm s} \phi(x)
\, , &
                         \label{eq:4.2}
\end{eqnarray} 
where $R_{\rm s}$ is the curvature of $h^{\alpha\beta}$. 
Imposing Weyl invariance at the 1-loop level 
to get rid of the ultraviolet divergences translates into the 
requirement that the so called beta-functions associated with the background 
fields vanish. The beta-function associated to the metric $g_{\mu\nu}$ is 
$\beta_{\mu\nu}^g = R_{\mu\nu} - \frac{1}{4} {H_\mu}^{\lambda\sigma}
H_{\nu\lambda\sigma}+ 2\nabla_\mu\nabla_\nu$ which should be set to zero. 
The 3-form $H$ is related to $B$ through 
$H_{\mu\nu\lambda} = \nabla_{\mu} B_{\nu\lambda}
+ \nabla_{\nu} B_{\lambda\mu} + \nabla_{\lambda} B_{\mu\nu}$. The other 
$\beta-$functions are 
$\beta_{\mu\nu}^B = \nabla_{\lambda} {H_{\mu\nu}}^{\lambda} 
- 2\left( \nabla_{\lambda}\phi{H^{\lambda}}_{\mu\nu}\right)=0\,$, 
$\beta^{\phi} = R + 2\Lambda + 4 \nabla^2\phi - 4 \left(\nabla\phi\right)^2 
-\frac{1}{12} H^2=0\,$. The constant $\Lambda$ is connected to the dimension 
of spacetime. For the bosonic string $\Lambda = \frac{D-26}{6\alpha'}$, whereas 
for the supersymmetric string with fermions $\Lambda\propto (D-10)$. The  
dimensions $D=26,10$ are the critical dimensions for the bosonic and 
supersymmetric strings, respectively, because in these dimensions the 
theory is free from divergences and anomalies. 
However, one can go away from these 
dimensions to the more familar 2, 3 or 4, by considering additional internal 
conformal field theories with central charges to complete, so to speak, 
the other extra dimensions. 

\vskip5mm
\centerline{\epsffile{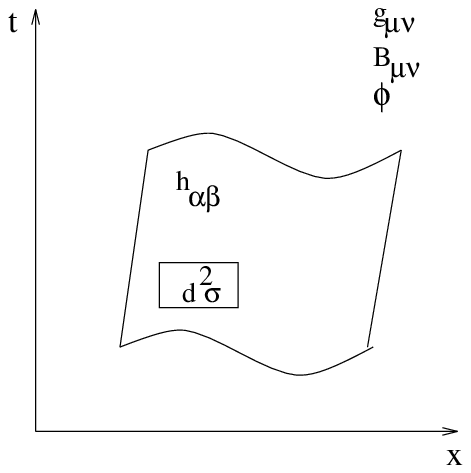}}
\vskip 3mm
{\noindent Figure 8. Spacetime diagram showing the nomenclature for the 
propagation of strings.}
\vskip 5mm

The equations for the three $\beta-$functions 
are the field equations of first order string theory,   
which can be derived from a spacetime effective action given by 
\begin{equation}
S_{\rm eff} = \frac{1}{4\pi}\int d^{D} x \sqrt{-g} e^{-2\phi} 
\left( R - 2\Lambda + 4 (\nabla\phi)^2
+ \frac{1}{12} H^2 \right).
                         \label{eq:4.3}
\end{equation} 
The Maxwell field $F_{\mu\nu}$  has been left out 
in this discussion (compare (\ref{eq:4.3}) and (\ref{eq:2.4})), 
as well as other fields like the tachyon $T$ of the bosonic 
string, but they  can be included consistently. 
Puting $D=2$ and $H=0$ in the equations of motion derived from (\ref{eq:4.3}) 
one finds a $2D$ BH solution in \cite{mandaletal91} given by
\begin{equation} 
ds^2= - (1-e^{-2\lambda r}) dt^2 + 
\frac{dr^2}{1-e^{-2\lambda r}} \,\,\, , e^{-2\phi}=e^{-\lambda r}\, ,
                         \label{eq:4.4}
\end{equation} 
where $\lambda^2\equiv-\frac{\Lambda}{2}$. This solution has horizons 
at $r_+=0$ and a singularity at $r=-\infty$. The Penrose diagram is identical 
to the Schwarzschild diagram in figure 1. 
Since this is a solution of the low-energy 
action it is only valid as long as the curvature is small compared to the 
Planck curvature. Is there a way to find an exact solution of the full 
action, i.e, of the world-sheet action, without resorting to perturbation 
theory? Yes, and the idea was initiated in \cite{witten91}.  One 
starts with the Wess-Zumino-Novikov-Witten (WZNW) model described by the 
action
\begin{equation}
S_{\rm WZNW} [g] = \frac{k}{8\pi} 
\int d^2\sigma \sqrt{h} h^{\alpha\beta} 
{\rm tr} \left( \nabla_\alpha g^{-1} \nabla_\beta g\right)
+ ik \Gamma(g) \, , 
                         \label{eq:4.5}
\end{equation} 
where $g$ is an element of some group, function of a field $x^\mu$, 
$k$ is a real and positive number (called the level of the Kac-Moody 
algebra) and the last term is the Wess-Zumino term which garantees 
conformal invariance of the action and for the purposes used here 
is of no importance.  The motivation for this model 
comes from the need to simplify the background in order to find solutions. 
One good simplification is to assume string propagation in a group 
manifold of a Lie group $G$ with elements $g$. Note the analogy of  
(\ref{eq:4.5}) with the world-sheet action  (\ref{eq:4.1}), 
where the trace has the role of a metric.
Now, if one supposes that $g\in SL(2,R)/U(1)$ one can parametrize it by
\begin{equation}
\left(
\begin{array}{clcr}
   a & u \\
  -v & b                        
\end{array}
\right)  
                \label{eq:4.6}  
\end{equation} 
with $ab+uv=1$. Since $SL(2,R)$ has dimension 3, and $U(1)$ has dimension 1, 
the quotient space group manifold $SL(2,R)/U(1)$ has dimension 2, 
which, in turn, can be parametrized by the coordinates $u,v$. 
After imposing that the action  (\ref{eq:4.5}) is gauge invariant and 
solving the equations of motion one finds \cite{witten91} 
\begin{equation}
S_{\rm WZNW} [g] = \frac{k}{4\pi} 
\int d^2\sigma \sqrt{h} h^{\alpha\beta} 
\frac{\nabla_\alpha u \nabla_\beta v}{1-uv} \, .
                         \label{eq:4.7}
\end{equation} 
Comparing with the world-sheet action (\ref{eq:4.1}) one immediatly finds 
that the target space metric is 
\begin{equation}
ds^2 = \frac{du dv}{1-uv} \, ,
                         \label{eq:4.8}
\end{equation} 
which upon further coordinate transformation  
can be put in the form (\ref{eq:4.4}). The dilaton can also be made 
to enter in this picture, see \cite{witten91}. Since one has to solve 
the classical equations of motion this treatment is semiclassical. 
The full treatment was attempted in \cite{dijkraffetal92} where it was 
found without approximations that the metric and dilaton are given by 
\begin{eqnarray}
& ds^2 = 2(k-1) \left[
-\left( \frac{x+1}{x-1} - \frac{2}{k}\right)^2 dt^2
+\frac{dx^2}{4(x^2-1)} \right] & \nonumber \\
& e^{-2\phi} = \frac{x^2-1}{\left(
\frac{x+1}{x-1} - \frac{2}{k} \right)^2} \, , &
                         \label{eq:4.9}
\end{eqnarray} 
where $x$ is a new radial coordinate. In the semiclassical approximation, 
when $k\rightarrow\infty$ one recovers Witten's result. The causal structure 
is given in figure 9 \cite{perryteo93}, the novel feature being that in the 
exact solution of the full theory the BH has no singularities! 
This indicates that string theory has indeed new things to show at 
the singularities. 

\vskip5mm
\centerline{\epsffile{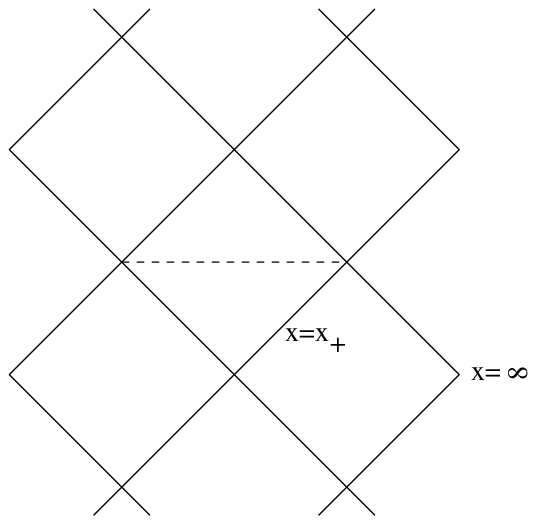}}
\vskip 3mm
\centerline{Figure 9. Penrose diagram for the non-singular 2D BH in 
string theory.}
\vskip 5mm

Having this exact solution and using the tools of string theory, 
namely, conformal field theory, one can in principle know how 
strings propagate in the BH background, calculate the latest 
stages of the BH evaporation and solve the information paradox. 
However, in practice the problem is still out of reach \cite{becker}.
Extensions to $4D$ of the idea of using a WZNW model to find exact 
solutions with associated conformal field theories have been tried 
with some interesting but limited progress \cite{johnsonmyers}.

We have just seen that the dilaton gives non-trivial dynamics to $2D$.  
This  has been known since the works of 
Teitelboim \cite{teitel82} and Jackiw \cite{jackiw82} where the 
power of $2D$ theories was first understood. They proposed the theory 
\begin{equation}
S = \frac{1}{2\pi} \int d^2x \sqrt{-g} e^{-2\phi}
\left( R- 2\Lambda \right) \, ,
                         \label{eq:4.10}
\end{equation}
with $\Lambda<0$. Although spacetime has constant and negative 
curvature it is possible to find a BH solution which is  
asymptotically ADS  \cite{lemossaMPLA,cadoni,ortizachuc}. 
The thermodynamics of this BH has been study (see this volume 
\cite{lemosthisvolume} and \cite{lemosthermodybhPRD}). 

In trying to find meaningful $2D$ actions one can look 
for connections with $4D$ general relativity, as it was done 
for $3D$ theories (see last section). Starting with 
the Einstein-Hilbert action 
$S = \frac{1}{16\pi}\int d^4x \sqrt{-g} (R-2\Lambda)$ 
and imposing planar symmetry (two-killing vectors), with 
a metric given by $ds^2 = g_{ab}dx^adx^b + e^{-2\phi} 
\left( dx^2 + dy^2\right)$, one finds upon dimensional reduction 
the following $2D$ action \cite{lemoscqg95}
\begin{equation}
S = \frac{1}{2\pi} \int d^2x \sqrt{-g} e^{-2\phi}
\left( R+ 2(\nabla\phi)^2 - 2\Lambda \right) \, . 
                         \label{eq:4.11}
\end{equation}
This theory also possesses a BH which, when reinterpreted in $4D$ 
yields a black membrane in general relativity \cite{lemoscqg95}. 
An obvious generalization of these three $2D$ theories is given 
by the Brans-Dicke action \cite{lemossaPRD1994}
\begin{equation}
S = \frac{1}{2\pi} \int d^2x \sqrt{-g} e^{-2\phi}
\left( R+ 4\omega(\nabla\phi)^2 - 2\Lambda \right) \, , 
                         \label{eq:4.12}
\end{equation}
where $\omega$ is a free parameter,  and  
$\omega=-1,-\frac12,0$ corresponding 
to string theory, planar general relativity and the Teitelboim-Jackiw 
theory, respectively.  When $\omega\rightarrow\infty$ one obtains the 
$2D$ analogue of general relativity \cite{lemossa94}, also called the 
$R=T$ theory \cite{mann}. The BH in this case is a massless BH as 
has been shown in \cite{lemossaPRD1994}. 
The BHs of action (\ref{eq:4.11}) for all rational $\omega$s 
have been analysed in detail in \cite{lemossaPRD1994} and the quantum 
version in \cite{hayward}. 
What about the temperature of these BHs? Usually the temperature goes 
with some power of the mass $M$, $T\propto M^{\gamma}$, where for instance 
for $\omega=0$, $\gamma=\frac12$ 
\cite{lemosthisvolume,lemosthermodybhPRD}. Thus, these 
$2D$ theories cannot tell much about the latest stages of the BH evaporation. 
A notable exception is string theory ($\omega=-1$) for which $\gamma=0$ and 
$T\propto {\rm constant}$, independent of the mass. Thus, following this 
result, the BH radiates indefinitely, which cannot be correct. In order 
to remedy the situation one has to make a full quantum treatment of 
the backreaction (see e.g. \cite{callanetal,hawking92}). 
For futher extensions and interests on lower dimensional BHs see, e.g., \cite{mannreview}.

\vskip 5mm

\noindent
{\bf 5. BHs in higher ${\bf D}$} 
\vskip 2mm

We have been considering BHs in general relativity, Brans-Dicke 
and string theories in 4 and lower dimensions. 
However, higher dimensional BHs are also important 
to study since they may shed some light 
on the understanding of non-perturbative effects in quantum
gravity (such as the compactification scheme), as well as 
expose which of the features of the usual 
four-dimensional BH solutions remain in higher dimensions.
Let us then go on to higher dimensions and consider,  
for a change, the original Kaluza-Klein theory in 5D. This is simply 5D 
general relativity in which the fifth dimension is a Killing direction, 
i.e., the fields are independent of the 5th dimension, $x^5$, say. 
The theory has two descriptions, the first given by the action 
\begin{equation}
S = \frac{1}{16\pi}\int d^5x \sqrt{-g} R \, ,
                         \label{eq:5.1}
\end{equation}
and metric components $g^{(5)}_{\mu\nu}$, $g^{(5)}_{\mu5}$ and 
$g^{(5)}_{55}$, $\mu,\nu=0,1,2,3$. In the other description the 
action takes the form
\begin{equation}
S = \frac{1}{16\pi} \int d^4x \sqrt{-g} 
\left(R - 2(\nabla\phi)^2 - e^{2\sqrt3 \phi} F^2 \right) \, , 
                         \label{eq:5.2}
\end{equation}
with the 5D metric related to the 4D fields by the usual Kaluza-Klein 
ansatz, 
$g^{(5)}_{\mu\nu}=e^{\frac{2\phi}{\sqrt3}}
\left(
g^{(4)}_{\mu\nu} + e^{-2\sqrt3 \phi} A_\mu A_\nu 
\right)$, $g^{(5)}_{\mu5}=e^{-\frac{4\phi}{\sqrt3}}A_\mu$, and 
$g_{55} = e^{-\frac{4\phi}{\sqrt3}}$. 
Due to this connection, one can generate with little effort static 
non-vacuum solutions from static vacuum solutions. Given a 
static vacuum 4D metric one can take its product with the real line 
$R$, 4D solution$\times R$, to obtain a 5D solution with two symmetry 
directions $(t,x^5)$. If one boosts this 5D solution in the 5th direction 
it still satisfies the 5D equations. However, when reinterpreted in 4D 
one obtains a solution with non-zero Maxwell and dilaton fields. 
In other words, given a 4D metric $g_{\mu\nu}$ one obtains 
a  new solution (${\tilde g}_{\mu\nu},{\tilde A}_\mu,{\tilde \phi}$) 
given by the transformations, 
\begin{eqnarray}
& {\tilde g}_{tt} = \frac{g_{tt}}{(\cosh^2\alpha + 
g_{tt}\sinh^2\alpha)^\frac12}\, , 
&  
\nonumber \\
& {\tilde g_{ij}} = g_{ij } {(\cosh^2\alpha + g_{tt}\sinh^2\alpha)^\frac12} 
\, ,
& 
\nonumber \\
& {\tilde A}_t = \frac{1+g_{tt}\sinh2\alpha}{2(\cosh^2\alpha + g_{tt}\sinh^2\alpha)} \, , &  
\nonumber \\
& e^{-\frac{4{\tilde\phi}}{\sqrt3}} = \cosh^2\alpha + g_{tt}\sinh^2\alpha\, 
                         \label{eq:5.3}
\end{eqnarray}
where $\alpha$ is the boost parameter and $i,j=1,2,3$. 
Example: given the Schwarzschild solution (\ref{eq:2.2}) one obtains after 
performing the above transformations, the following 
\cite{dobiasch,chodos,wiltshire}
\begin{eqnarray}
& ds^2 = - \frac{1-\frac{r_+}{r}}{\sqrt{1-\frac{r_-}{r}}} dt^2 +
\frac{dr^2}{\left(1-\frac{r_+}{r}\right)\left(1-\frac{r_-}{r}\right)} + 
r^2 \left(1-\frac{r_-}{r}\right) d\Omega_2^2 \, ,
&  
\nonumber \\
& A_t = \frac{\sqrt{r_+r_-}}{r} \,\,\,, \,\,\,\,  
e^{-\frac{4\phi}{\sqrt3}} = 1 - \frac{r_-}{r} \, ,  & 
                         \label{eq:5.4}
\end{eqnarray} 
where we have redifined the Schwarzschild radial coordinate  ($r_S$, say) 
in (\ref{eq:2.2}) to $r_S = r\left(1-\frac{r_-}{r}\right)$, 
and put $r_-=2m\sinh^2\alpha$, $r_+=2m\cosh^2\alpha$, $m$ being 
the Schwarzschild mass. The ADM mass and electric charge 
are $M=m\sinh2\alpha$, $Q=m\cosh2\alpha$, respectively. There 
are horizons at $r=r_\pm$ and the singularity is at $r=0$.  
Another type of transformation, called Harrison transformation 
\cite{harrison}, transforms metrics within general relativity, taking 
for instance, the Schwarzschild metric into the Reissner-Nordstrom metric. 
Now, in string theory there is the analogue of these boost transformed 
solutions. In a simple case, one starts with a static solution 
(${g}_{\mu\nu},\phi$), with $B_{\mu\nu}=0$ and $A_{\mu}=0$. Then one gets 
a new solution (${\tilde g}_{\mu\nu},{\tilde A}_\mu,{\tilde \phi}$) 
by making the following transformations \cite{hassan92}
\begin{eqnarray}
& {\tilde g}_{tt} = \frac{g_{tt}}{(\cosh^2\alpha + 
g_{tt}\sinh^2\alpha)^2}\, , 
&  
\nonumber \\
& {\tilde A}_t = 
\frac{1+g_{tt}\sinh2\alpha}{2\sqrt2(\cosh^2\alpha + 
g_{tt}\sinh^2\alpha)} \, , &  
\nonumber \\
& e^{-2{\tilde\phi}} = e^{-2\phi}\cosh^2\alpha + g_{tt}\sinh^2\alpha\, . 
                         \label{eq:5.5}
\end{eqnarray}
Recalling that the Schwarzschild solution (\ref{eq:2.2}) is a solution 
of string theory, one can apply  (\ref{eq:5.5}) to obtain the electric 
charged BHs given in equation (\ref{eq:2.78}). But we are still 
discussing $4D$ BHs.

To obtain charged BHs in higher $D$, one starts with a $D$-dimensional 
uncharged BH \cite{tangherlini}, 
\begin{equation}
ds^2 = - (1-\frac{cm}{r^n}) dt^2 + \frac{dr^2}{1-\frac{cm}{r^n}} 
+ r^2 d\Omega_{n+1}^2 \, ,
                         \label{eq:5.6}
\end{equation}
where $n=D-3$ and c is a constant. 
This is a solution of both $D-$dimensional general relativity 
and string theory. By using the transforming equations 
(\ref{eq:5.5})  one can obtain 
the $D-$dimensional electrically charged BHs in string theory
\cite{maeda88}, 
\begin{eqnarray}
& ds^2 = -\left(1-\frac{cm}{r^n}\right) 
\left(1+\frac{cm\sinh^2\alpha}{r^n}\right) dt^2 
+ \frac{dr^2}{1-\frac{cm}{r^n}} + r^2 d\Omega_{n+1}^2 \, ,
&  
\nonumber \\
& A_{t} = - 
\frac{cm\sinh2\alpha}{2\sqrt2\left(r^n+cm\sinh^2\alpha\right)} \, ,
& 
\nonumber \\
& e^{-2\phi} = 1 + \frac{cm}{r^n} \sinh^2\alpha \, .
                         \label{eq:5.7}
\end{eqnarray}
The ADM mass and charge are given by 
$M=m(1+\frac{2n}{n+1}\sinh^2\alpha)$ and 
$Q=\frac{c\, m\, n\, \sinh2\alpha}{\sqrt2}$. The event horizons are 
at $r=(c\, m)^{\frac1n}$, and the singularities at $r=0$. In constrast 
with $4D$ we have that in the extremal limit the singularity is timelike 
rather than null, and the temperature of the extreme BH is zero. 
There are no higher $D$ magnetically charged BHs because there are no Maxwell 
magnetic charges (one cannot integrate a 2-form $F$ over a $D-2$ sphere).
However, using a magnetic charge associated with the 3-form field $H$, 
one can find magnetically charged BH solutions in string theory 
\cite{horowitzstrominger}. 

From BHs in $D-$dimensions one can find straightforwardly black strings in 
$(D+1)-$dimensions. It is only necessary to take the product of the BH 
with $R$ \cite{horowitzstrominger}, 
\begin{equation}
ds^2 = - (1-\frac{cm}{r^n}) dt^2 + \frac{dr^2}{1-\frac{cm}{r^n}} 
+ r^2 d\Omega_{n+1}^2 + dx^2\, .
                         \label{eq:5.8}
\end{equation} 
If one takes the product of the BH with $R^2$, $R^3$, $R^p$, one obtains 
a black membrane, a black 3-brane, and a black p-brane. These branes 
are simple products. For instance, to get a black string  that is not 
a simple product one performs, after Lorentz boosting to get charge, 
a T-duality transformation on the simple product black string 
to obtain 
\begin{eqnarray}
& ds^2 = - \frac{ \left(1-\frac{cm}{r^n}\right) }
{ \left(1+\frac{cm\sinh^2\alpha}{r^n}\right) } dt^2 
+ \frac{dr^2}{1-\frac{cm}{r^n}} + r^2 d\Omega_{n+1}^2 
+ \frac{dx^2}{1+\frac{c\, m\, \sinh^2\alpha}{r^n}} \, ,
&  
\nonumber \\
& B_{xt} = - 
\frac{cm\sinh2\alpha}{2\left(r^n+c\, m\, \sinh^2\alpha\right)} \, ,
& 
\nonumber \\
& e^{-2\phi} = 1 + \frac{c\, m}{r^n} \sinh^2\alpha \, .
                         \label{eq:5.9}
\end{eqnarray}
The causal structure is identical to Schwarzschild. In the extremal limit 
the metric field is given by 
\begin{equation}
ds^2 = \frac{-dt^2 + dx^2}{1+\frac{{\overline c}\, m}{r^n}} 
+ dr^2 + r^2 d\Omega_{n+1}^2 \, , 
                         \label{eq:5.10}
\end{equation}
where ${\overline c}$ is a redefinition of $c$. There are 
two novel features in this solution (\ref{eq:5.10}): 
(i) an extra symmetry has appeared, the metric is now 
boost-invariant in the $(x,t)$ plane, and (ii) 
the solution is the same solution found in \cite{dabholkharetal91} 
for a straight fundamental macroscopic string. These objects appear as 
stable extended sates of closed-string theories and are distinct from 
the cosmic strings of string theory. This means that fundamental strings 
are extreme black strings. There is no such analogue  in general 
relativity. The electron, a fundamental particle is not an extreme BH.

Ultimately, one would like to get a BH solution of $10D$ string theory, 
suitably dimensionally reduced to $4D$. One starts with the 
$10D$ action
\begin{equation}
S=\frac{1}{16\pi}\int d^{10} x \sqrt{-G} \left[ 
R_G + \nabla_M\Phi \nabla^M\Phi 
- \frac{1}{12} H^2 
- \frac14 {F^I}^2 \right]
                         \label{eq:5.11}
\end{equation}
where $H^2 = H_{MNP} H^{MNP}$, ${F^I}^2 = F_{MN}^I F^{I{MN}}$, 
capital letters denote $10D$ fields and indices, and $I$ is 
an internal index.  
Through a Kaluza-Klein reduction to $4D$, one can find an effective 
$4D$ action, with the other dimensions compactified on a six torus. 
One writes the ansatz, 
\begin{eqnarray}
{G_{MN}} = 
\left(
\begin{array}{clcr}
  e^{2\phi}g_{\mu\nu} + G_{mn}A_{\mu}^m A_{nu}^n & A_\mu^m G_{mn} \\
  A_\nu^n G_{mn}                                 & G_{mn}
                        \end{array}
\right)  
                           \label{eq:5.12}
\end{eqnarray}
with the $4D$ spacetime indices $\mu\, \nu = 0,1,2,3$, 
$m\, , n=1,...,6$, and $\phi$ and $A$ are the $4D$ dilaton and Kaluza-Klein 
$U(1)$ fields, respectively. The action (\ref{eq:5.10}) then turns into 
\begin{eqnarray}
& S = \frac{1}{16\pi}
\int d^4x \sqrt{-g} \left( R - \frac12 \nabla_\mu\phi \nabla^\mu\phi
- \frac{1}{2}  e^{2\phi} \nabla_\mu\psi \nabla^\mu\psi
 \right. 
&
\nonumber \\
& 
\left. 
- \frac{1}{4} e^{-\phi} F_{\mu\nu} F^{\mu\nu}
+ \frac{1}{8} {\rm tr} ( \nabla_\mu {M} 
\,  \nabla^\mu 
{M}\,  ) \right) \, ,
                         \label{eq:5.13}
\end{eqnarray}
where $M$ is a $O(6,22)$ matrix of the scalar (moduli) fields 
appearing in the reduction process  and $\psi$ is the axion 
related to $H_{\mu\nu\lambda}$ by 
$H_{\mu\nu\lambda} = \frac{e^{2\phi}}{\sqrt{-g}} 
\epsilon^{\mu\nu\lambda\rho}\nabla_\rho \psi$, see 
\cite{cveticyoum} for all details.  
This is quite complicated to solve, but applying a generalized 
boosting procedure and using all the symmetries it is possible 
to find the most general BH solution with all charges \cite{cveticyoum}. 
An important consequence brought from this 4D analysis is that 
the extreme BH solutions correspond to massive excitations of 4D 
superstrings, suggesting that BHs are simple string states 
\cite{duff94} 
and confirming the idea that elementary particles (represented here 
by those string states) might behave like BHs. These BHs saturate 
the Bogomolniy-Prasad-Somerfield bound of the underlying supersymmetric 
theory and are called extreme BPS BHs. 

There are also studies on black p-branes in string theory  
(e.g. \cite{duff,maldacena})
motivated by their importance in the non-perturbative dynamics of the 
$11D$ M--theory \cite{townsend96}, a theory not explicitly formulated, 
but  known to agglutinate the four consistent 
(heterotic, type I, type IIA and B) superstring theories.

We have been presenting higher dimensional BH solutions in 
Kaluza-Klein theory, string theory and general relativity.  
Yet, although pure general relativity can be 
formulated in other dimensions, when one goes to dimensions higher than 
four it is not anymore unique. The natural generalization is given by the 
Lovelock action  \cite{lovie} so that the field equations for the metric 
remain of second order. 
The theory can also be considered as a dimensional continuation of 
the Euler densities of lower dimensions \cite{regge,zumino,TZ}. 
In four dimensions 
one has to take in consideration two Euler densities. The Euler density 
of the 0-dimensional space which is proportional to $\sqrt{-g}$, 
and the Euler density of the 2-dimensional space, proportional to 
$\sqrt{-g} R$, where $g$ is the determinant of the metric and $R$ the 
Ricci curvature scalar. Thus Lovelock gravity in four dimensions reduces 
to Einstein gravity, with action 
$\frac{1}{16\pi G}\int d^4 x \sqrt{-g}(-2\Lambda+R)$.  
A similar construction and 
action is obtained for three dimensions. In six dimensions one 
has still to add the Euler characteristic of four dimensional space, 
i.e. the Gauss-Bonnet term, to have the Lanczos action, given by,  
$\frac{1}{16\pi G}\int d^6 x\sqrt{-g}\left(-2\Lambda+R+
\alpha_2(R_{\alpha\beta\gamma\sigma}R^{\alpha\beta\gamma\sigma}
-4R_{\alpha\beta}R^{\alpha\beta}+R^2)\right)$, where $\alpha_2$ 
is a new constant. A similar construction and action can be obtained for 
five dimensions. 
For each two new dimensions there exists a new constant $\alpha_p$. 
These constants do not seem to have a direct physical meaning.  In 
order to find a meaningful set of constants in any dimension $D$, 
it was proposed in \cite{giambiagi,BTZ} a method which 
restricts drasticaly the number of independent constants to two, 
$G$ and $\Lambda$, thus yielding a restricted Lovelock gravity. 
This method separates, in a natural manner, 
theories in even dimensions ($D=2n$, with $n=1,2,..$) 
from theories in odd dimensions  ($D=2n+1$).
The BH solutions are given by \cite{BTZ}
\begin{eqnarray}
& ds^2 = - 
\left[ 1 - \left(\frac{2sM}{r^p}+q\right)^{\frac{1}{n-1}} + (\frac{r}{l})^2 \right] dt^2 
+ \frac{dr^2}{ 1 - 
\left(\frac{2sM}{r^p}+q\right)^{\frac{1}{n-1}} + (\frac{r}{l})^2} &
\nonumber \\
&
+r^2 d\Omega_{D-2}^2 \, , &  
                         \label{eq:5.14}
\end{eqnarray}
where for odd $D$ one puts ($s=\frac12,p=0,q=1$), and for even $D$ 
one has ($s=1,p=1,q=0$). 
There are horizons at $r=r_+$ given by the zeros of $g^{rr}$ and the 
singularity is at $r=0$. Note that there is no restriction in the 
dimension of spacetime, it can be any natural number from $3$ to $\infty$. 
Since in general relativity BHs appear as the final 
state of gravitational collapse it is important to know if the BH 
solutions found in Lovelock gravity can, in an analogous manner, 
form from gravitational collapse. It was shown that, indeed, Lovelock 
BHs form from regular initial data \cite{ilhalemos0,ilhalemos1}.  The 
collapsing matter is modelled by a Friedmann type metric, and  the 
solution can be viewed as a dimensional continued Oppenheimer-Snyder 
gravitational collapse. A possible scenario for the occurrence of this 
collapse in $D$ dimensions, would be in the very early universe, 
before the $(D-4)$ extra dimensions have been compactified. In 
turn, these newly formed higher dimensional BHs could play a 
role in the compactification process. 
It is interesting to note that these BH and 
collapsing solutions  show that some 
important features of classical general relativity are preserved and 
carried into Lovelock gravity in any dimension.

\vskip 4mm

\noindent
{\bf 6. Conclusions} 
\vskip 1mm

We have investigated BH, black string and black membrane solutions in 
several dimensions and in several theories (general relativity, Kaluza-Klein, 
Brans-Dicke, Lovelock gravity and string theory).  We 
have seen that new properties come into play. For instance, in string 
theory there are BHs without singularities.  
It was also shown that the existence of a negative 
cosmological term can be important in producing black solutions, as was
the case of black strings in $4D$ general relativity.  We have also 
seen that some features appearing in general relativity remain in other 
theories, like in Lovelock gravity, where the BHs also form from  
gravitational collapse of matter. Other important developments not 
discussed here are solutions of BHs with both electric and magnetic 
charges, rotating BHs in several $D$, 
duality between charge and angular momentum, 
and multi-BH solutions in the various theories, to name a few. 
 
With such a profusion of BHs in all these gravity theories, one 
could hope to understand in some detail the BH evaporation process, 
at least, in one of those solutions. 
However, the problem of calculating Hawking radiation of BHs, black 
strings, black membranes or black p-branes, 
through the latest stages of the evaporation process, remains. 

A remarkable property of BHs is that they appear in all scales, from 
the Planck length to astronomical dimensions. This seems to be unique. 
Electrons, molecules, stars and galaxies have well 
defined scales,  BHs do not. 
\vfill\eject
\noindent {\bf Acknowledgements --} I thank Paulo S\'a and Vilson 
Zanchin for collaborations and conversations. I also thank Antares 
Kleber for reading the manuscript carefully.


\vskip 2mm
\begin{thebibliography}{100}

\bibitem{oppie}  J. R. Oppenheimer, H. Snyder, {\it Phys. Rev.} 
{\bf 56}, 455 (1939). 
\bibitem{zwickybaade}  W. Baade, F. Zwicky, {\it Phys. Rev.} {\bf 45},
138 (1934).
\bibitem{einstein39} A. Einstein, {\it Ann. Math. (Princeton)} {\bf 40}, 
 922, (1939). 
\bibitem{michel1784} J. Michell, {\it Phil. Trans. R. Soc. (London)} {\bf 74}, 
35 (1784).
\bibitem{laplace1796} P. S. Laplace, {\it Exposition du Syst\`eme du 
Monde} (J. B. M. Duprat, Paris, 1796). 
\bibitem{penrose69} R. Penrose, {\it Riv. Nuovo Cimento} 
{\bf 1}, 252 (1969).
\bibitem{lemosprl92} J. P. S. Lemos, {\it Phys. Rev. Lett.}, {\bf 68}, 
1447  (1992). 
\bibitem{joshi93} P. S. Joshi, {\it Global Aspects in Gravitation 
and Cosmology}, (Clarendon Press, Oxford, 1993). 
\bibitem{lemos96} J. P. S. Lemos, in {\it Proceedings of the 
XXI$^{{\underline{th}}}$ Annual Meeting of the Sociedade 
Astron\^omica Brasileira
(August 1995)}, eds. F. Jablonski, F. Elizalde, L. Sodr\'e Jr., 
V. Jatenco-Pereire, (IAG 1996), p. 57.
\bibitem{lynden-bell}  D. Lynden-bell,  {\it Nature} {\bf 223}, 690 (1969). 
\bibitem{penrose65} R. Penrose, {\it Phys. Rev. Lett.} {\bf 14} (1965) 57. 
\bibitem{hawking74} S. W. Hawking, {\it Nature} {\bf 248}, 30 (1974). 
\bibitem{hawking76} S. W. Hawking, {\it Phys. Rev. D} {\bf 13}, 191 (1976). 
\bibitem{strominger96} A. Strominger, {\it Phys. Rev. Lett.} {\bf 77} (1996) 
3498. 
\bibitem{colemanetal1992} S. Coleman, J. Preskill, F. Wilczeck, 
{\it Nucl. Phys.} {\bf B378}, 175 (1992). 
\bibitem{carter74} B. Carter, {\it Phys. Rev. Lett.} {\bf 33}, 558 (1974). 
\bibitem{gibbonshull1982} G. Gibbons, C. M. Hull, {\it Phys. Lett.} 
{\bf B109}, 190 (1982). 
\bibitem{hawking1970} S. W. Hawking,  {\it Mon. Not. R. astr. Soc.} 
{\bf 152}, 75 (1971). 
\bibitem{townsend96} P. K. Towsend, {\it M-theory for mortals},  
Lectures delivered at the XVII$^{ {\underline{\rm th}} }$ UK Institute for 
Theoretical High Energy Physicists (1996). 
\bibitem{witten95} E.  Witten, {\it Nucl. Phys.} {\bf B443}, 85 (1995).
\bibitem{vafa} C. Vafa, hep-th/9602022.  
\bibitem{greenschwarzwitten} M. B. Green, J. H. Schwarz, E. Witten, 
{\it Superstring theory}, (Cambridge Unversity Press, Cambridge 1987). 
\bibitem{tangherlini} F. R. Tangherlini, {\it Il Nuovo Cim.} {\bf XXVII}, 
636 (1963). 
\bibitem{horowitzreview} G. Horowitz, {\it Proceedings of The 1992 Trieste 
Spring School on String Theory and Quantum Gravity}, (World Scientific, 
Singapore 1993), hep-th/921019. 
\bibitem{carlipCQG} S. Carlip, {\it Class. Quantum Grav.} 
{\bf 12}, 2853 (1995).   
\bibitem{hawking75} S. W. Hawking, {\it Comm. Math. Phys.} {\bf 43},  
149 (1975).  
\bibitem{hawking79}  S. W. Hawking, {\it in General Relativity}, 
eds. S. W. Hawking, W. Israel (Cambridge University Press, 
Cambridge 1979). 
\bibitem{lemosplb95} J. P. S. Lemos, {\it Phys. Lett. B} {\bf 352}, 46 (1995).
\bibitem{lemoscqg95}  J. P. S. Lemos, {\it Class. Quantum Gravity} {\bf 12}, 1081 (1995). 
\bibitem{crusciel} P. T. Chrusciel, {\it Contemporary Mathematics -- AMS} 
{\bf 170}, 23 (1994). 
\bibitem{gibbons82} G. Gibbons, {\it Nucl. Phys.} {\bf B207}, 337 (1982).
\bibitem{maeda88} G. Gibbons, K. Maeda, {\it Nucl. Phys.} {\bf B298}, 
741 (1988).
\bibitem{garfinkle91} D. Garfinkle, G. Horowitz, A. Strominger, 
{\it Phys. Rev. D} {\bf 43}, 3140 (1991); {\bf 45}, 3888(E) (1992). 
\bibitem{sen92} A. Sen, {\it Phys. Rev. Lett.} {\bf 69}, (1992). 
\bibitem{Deseretal82} S. Deser, R. Jackiw, G. 't Hooft, 
{\it Ann. Phys.} {\bf 152}, 220 (1984).
\bibitem{achutownsen88} A. Ach\'ucarro, P. K. Townsend, {\it Phys. Lett.} 
{\bf B180}, 89 (1988). 
\bibitem{witten88} E. Witten, {\it Nucl. Phys.} {\bf B311}, 46 (1988). 
\bibitem{banadosetal92} M. Ba\~nados, C. Teitelboim and J. Zanelli, 
{\it Phys. Rev.Lett.} {\bf 69}, 1849 (1992).
\bibitem{bhtz93} M. Ba\~nados, M. Henneaux, C. Teitelboim and J. Zanelli, 
{\it Phys. Rev. D} {\bf 48}, 1506 (1993).
\bibitem{henneauxcousaert93} O. Coussaert, M. Henneaux, 
{\it Phys. Rev. Lett.} {\bf 72}, 183 (1994). 
\bibitem{mannross}  R. B. Mann, S. F. Ross, {\it Phys. Rev. D} {\bf 47}, 
3319 (1993).
\bibitem{horowitzwelch93} G. T. Horowitz, D. L. Welch, 
{\it Phys. Rev. Lett.} {\bf 71}, 328 (1993).
\bibitem{kaloper93} N. Kaloper, {\it Phys. Rev. D} {\bf 48} (1993) 2598.
\bibitem{kaloper94} N. Kaloper, {\it Phys. Rev. D} {\bf 48} (1993) 4658. 
\bibitem{lemoszanchin961} J. P. S. Lemos, V. T. Zanchin, 
{\it Phys. Rev. D} {\bf 53}, 4684 (1996). 
\bibitem{lemoszanchin962} J. P. S. Lemos, V. T. Zanchin, 
{\it Phys. Rev. D} {\bf 54}, 3840 (1996). 
\bibitem{hawkingellis} S. W. Hawking, G. F. R. Ellis, 
{\it The Large Scale Structure of Space-Time}, 
(Cambridge University Press, Cambridge, 1973).
\bibitem{thorne72} K. S. Thorne, in {\it  Magic without Magic}, ed. J. R.  
Klauder,  (Freeman and Company, San Francisco 1972), p. 231.
\bibitem{lemosmoniz97} J. P. S. Lemos, P. V. Moniz, ``Supersymmetry of 
the black strings'', in preparation. 
\bibitem{sakleberlemos96} P. M. S\'a, A. Kleber, J. P. S. Lemos, 
{\it Class. Quantum Grav.} {\bf 13}, 125 (1996).
\bibitem{salemos95} P. M. S\'a, J. P. S. Lemos, hep-th/9611169. 
\bibitem{kchan} K. C. Chan, R. B. Mann, {\it Phys. Rev. D} {\bf 50}, 
6385 (1994). 
\bibitem{mandaletal91} G. Mandal, A. M. Sengupta, S. R. Wadia, 
{\it Mod. Phys. Lett. A} {\bf 6}, 1685 (1991). 
\bibitem{witten91} E. Witten, {\it Phys. Rev. D} {\bf 44}, 314 (1991).
\bibitem{dijkraffetal92} R. Dijkgraaf, H. Verlinde, E. Verlinde,  
{\it NUcl. Phys.} {\bf B371}, 269 (1992). 
\bibitem{perryteo93} M. J. Perry, E. Teo, 
{\it Phys. Rev. Lett.} {\bf 70}, 2669 (1993). 
\bibitem{becker} K. Becker, {\it Strings, Black Holes and Conformal 
Field Theory}, (PhD thesis, University of Bonn 1994), hep-th/9404157. 
\bibitem{johnsonmyers} C. V. Johnson, R. C. Myers, 
{\it Phys. Rev. D} {\bf 52}, 2294 (1995).
\bibitem{teitel82}  C. Teitelboim, in {\it Quantum Theory of Gravity}, 
ed. S. M. Christensen (Hilger, Bristol, 1984). 
\bibitem{jackiw82} R. Jackiw,  in {\it Quantum Theory of Gravity}, 
ed. S. M. Christensen (Hilger, Bristol, 1984). 
\bibitem{lemossaMPLA} J. P. S. Lemos, P. M. S\'a, 
{\it Mod. Phys. Lett. A} {\bf 9}, 771 (1994). 
\bibitem{cadoni} M. Cadoni, S. Mignemi, {\it Phys. Rev. D} 
{\bf 51}, 4139 (1995).
\bibitem{ortizachuc}  A. Ach\'ucarro, M. E. Ortiz, 
{\it Phys. Rev. D} {\bf 48}, 3600 (1993). 
\bibitem{lemosthisvolume} J. P. S. Lemos, 
``Comparative study between the thermodynamics of the 2-dimensional  
black hole in the Teitelboim-Jackiw theory and the 4-dimensional 
Schwarzschild black hole'', this volume.  
\bibitem{lemosthermodybhPRD} J. P. S. Lemos, {\it Phys. Rev. D} {\bf 54}, 
6206 (1996). 
\bibitem{lemossaPRD1994} J. P. S. Lemos, P. M.  S\'a, 
{\it Phys. Rev. D.} {\bf 49}, 2897 (1994). 
\bibitem{lemossa94} J. P. S. Lemos, Paulo S\'a, 
{\it Class. Quantum Gravity} {\bf 11}, L11 (1994). 
\bibitem{mann} R. B. Mann, S. F. Ross, {\it Phys. Rev. D} {\bf 47}, 
3312 (1993).
\bibitem{hayward} J. D. Hayward, hep-th/9508090. 
\bibitem{callanetal} C. G. Callan, S. B. Giddings, J. A. Harvey, 
A. Strominger, {\it Phys. Rev. D} {\bf 45}, R1005 (1992).
\bibitem{hawking92} S. W. Hawking, {\it Phys. Rev. Lett.} {\bf 69}, 406 (1992). 
\bibitem{mannreview} R. B. Mann, gr-qc/9503024. 
\bibitem{dobiasch} P. Dobiasch, D. Maison, {\it Gen. Rel. Grav.} 
{\bf 14}, 231 (1982). 
\bibitem{chodos}A. Chodos, S. Detweiler, {\it Gen. Rel. Grav.} 
{\bf 14}, 870 (1982)
\bibitem{wiltshire} G. Gibbons, D. Wiltshire, {\it Ann. Phys.} {\bf 167}, 
201 (1986). 
\bibitem{harrison} B. Harrison, {\it J. Math. Phys.} {\bf 9}, 1744 (1968). 
\bibitem{hassan92} S. Hassan, A. Sen, {\it Nucl. Phys.} {\bf B375}, 
103 (1992). 
\bibitem{horowitzstrominger} G. Horowitz, A. Strominger, 
{\it Nucl. Phys.} {\bf B360}, 197 (1991). 
\bibitem{dabholkharetal91} A. Dabholkar, G. Gibbons, J. A. Harvey, 
F. Ruiz Ruiz, {\it Nucl. Phys.} {\bf B340}, 33 (1990). 
\bibitem{cveticyoum} M. Cvetic, D. Youm, 
{\it Nucl. Phys.} {\bf B472}, 249 (1996). 
\bibitem{duff94} M. J. Duff, R. R. Khuri, R. Minasian, J. Rahmfeld, 
{\it Nucl. Phys.} {\bf B418}, 195 (1994).  
\bibitem{duff} M. J. Duff, hep-th/9611203. 
\bibitem{maldacena} J. M. Maldacena, {\it Black Holes in String Theory}, 
(PhD thesis, University of Princeton 1996), hep-th/9607235. 
\bibitem{lovie}  D. Lovelock, {\it J. Math. Phys.} {\bf 12}, 498 (1971). 
\bibitem{regge} T. Regge, {\it Phys. Rep.} {\bf 137}, 31 (1986).
\bibitem{zumino} B. Zumino, {\it Phys. Rep.} {\bf 137}, 109 (1986).
\bibitem{TZ}  C. Teitelboim, J. Zanelli, in {\it Constraint Theory 
and Relativistic Dynamics}, eds. G. Longhi, L. Lussana, 
(World Scientific, Singapore 1987).
\bibitem{giambiagi}  M. Ba\~nados, C. Teitelboim, J. Zanelli, in 
{\it J. J. Giambiagi Festschrift}, eds.  H. Falomir et al 
(World Scientific, Singapore 1991). 
\bibitem{BTZ}  M. Ba\~nados, C. Teitelboim, J. Zanelli,
{\it Phys. Rev.} D {\bf 49}, 975 (1994). 
\bibitem{ilhalemos0} A. Ilha, J. P. S. Lemos, this volume. 
\bibitem{ilhalemos1} A. Ilha, J. P. S. Lemos,  
``Dimensionally continued Oppenheimer-Snyder 
gravitational collapse. solutions in even dimensions'', 
{\it Phys. Rev. D}, to appear (1997), hep-th/9608004. 





\end{thebibliography}
\end{document}